\documentclass[showpacs,twocolumn,preprintnumbers,amsmath,amssymb]{revtex4}

\usepackage{graphicx}
\usepackage{dcolumn}
\usepackage{bm}
\usepackage{multirow}
\usepackage{ulem}

\newcommand{\bq}   {\begin{equation}}
\newcommand{\eq}   {\end{equation}}
\newcommand{\bqa}  {\begin{eqnarray}}
\newcommand{\eqa}  {\end{eqnarray}}
\newcommand{\nn}   {\nonumber \\}

\def\be         {\begin{equation}}
\def\ee         {\end{equation}}
\def\bea        {\begin{eqnarray}}
\def\eea        {\end{eqnarray}}
\def\bnn        {\begin{eqnarray*}}
\def\enn        {\end{eqnarray*}}

\begin{document}

\title{A topological Fermi-liquid theory for interacting Weyl metals with time reversal symmetry breaking}
\author{Yong-Soo Jho, Jae-Ho Han, and Ki-Seok Kim}
\affiliation{Department of Physics, POSTECH, Pohang, Gyeongbuk 790-784, Korea}
\date{\today}

\date{\today}

\begin{abstract}
Introducing both Berry curvature and chiral anomaly into Landau's Fermi-liquid theory, we construct a topological Fermi-liquid theory, applicable to interacting Weyl metals in the absence of time reversal symmetry. Following the Landau's Fermi-liquid theory, we obtain an effective free-energy functional in terms of the density field of chiral fermions, where the band structure is modified, involved with an emergent magnetic dipole moment due to the Berry curvature. The density field of chiral fermions is determined by a self-consistent equation, minimizing the effective free-energy functional with respect to the order-parameter field. Beyond these thermodynamic properties, we construct Boltzmann transport theory to encode both the Berry curvature and the chiral anomaly in the presence of forward scattering of a Fermi-liquid state, essential for understanding dynamic correlations in interacting Weyl metals. This generalizes the Boltzmann transport theory for the Landau's Fermi-liquid state in the respect of incorporating the topological structure and extends that for noninteracting Weyl metals in the sense of introducing the forward scattering. Finally, we justify this topological Fermi-liquid theory, generalizing the first-quantization description for noninteracting Weyl metals into the second-quantization representation for interacting Weyl metals. First, we derive a topological Fermi-gas theory, integrating over high-energy electronic degrees of freedom deep inside a pair of chiral Fermi surfaces. As a result, we reproduce a topologically modified Drude model with both the Berry curvature and the chiral anomaly, given by the first-quantization description. Second, we take into account interactions between such low-energy chiral fermions on the pair of chiral Fermi surfaces. Following the Landau's Fermi-liquid theory, we perform the renormalization group analysis. We find that only forward scattering turns out to be marginal above possible superconducting transition temperatures, justifying the topological Fermi-liquid theory of interacting Weyl metals with time reversal symmetry breaking. The topological Fermi-liquid theory serves a theoretical platform for us to investigate the role of Fermi-liquid interactions in anomalous transport phenomena of interacting Weyl metals such as anomalous Hall effects, chiral magnetic and vortical effects, and negative longitudinal magnetoresistivity properties. In addition, it allows us to study how thermodynamic properties such as the Wilson's ratio and spectra of collective excitations such as zero sound modes in the Landau's Fermi-liquid state are modified due to the Berry curvature and the chiral anomaly.
\end{abstract}


\maketitle

\section{Introduction}

Weyl metals \cite{WM1,WM2,WM3} are described by pairs of Weyl bands, separated in momentum space, where each Weyl band describes emergent relativistic Weyl electrons. Although the band structure itself may be regarded to be a three dimensional version of a graphene, focusing on one pair of Weyl bands, the three dimensional character allows topologically identified nontrivial properties in Weyl metals \cite{WM_Reviews}. A pair of Weyl points can be identified with a magnetic monopole and anti-monopole pair in momentum space. Accordingly, the Berry curvature is assigned by this monopole pair, which turns out to play an essential role in anomalous transport phenomena of Weyl metals \cite{NLMR_First_Exp,NLMR_Followup_I,NLMR_Followup_II,NLMR_Followup_III,Nielsen_Ninomiya_NLMR,CME1,CME2,CME3,CME4,CME5,CME6,CME7,Boltzmann_Chiral_Anomaly1,Boltzmann_Chiral_Anomaly2,
Boltzmann_Chiral_Anomaly3,Boltzmann_Chiral_Anomaly4,Boltzmann_Chiral_Anomaly5,Boltzmann_Chiral_Anomaly6,AHE1,AHE2,AHE3,Disordered_Weyl_Metal1,Disordered_Weyl_Metal2},  where anomalous Hall effects, chiral magnetic and vortical effects, and negative longitudinal magnetoresistivity properties have been discussed. In particular, the negative longitudinal magnetoresistivity has been measured experimentally, proposed to be a fingerprint of a Weyl metal phase \cite{NLMR_First_Exp,NLMR_Followup_I,NLMR_Followup_II,NLMR_Followup_III}. However, these anomalous transport phenomena have been examined without taking into account interaction effects. If electron correlations are introduced into Weyl metals, possible interplays between the topological structure and interactions would modify such transport properties. 

The Wilson's ratio remains unchanged in the Landau's Fermi-liquid state, compared with a Fermi-gas phase \cite{Boltzmann_LFL}. Interaction effects on the specific heat coefficient give essentially the same renormalization as those on the uniform spin susceptibility, thus canceled in the Wilson's ratio, which leads to a ``weakly universal" constant against interactions. However, if both the Berry curvature and the chiral anomaly are introduced into the Landau's Fermi-liquid state, the Wilson's ratio would be modified. The zero sound mode may be regarded to be a fingerprint of the Landau's Fermi-liquid state, where the density-density correlation function in the random phase approximation (RPA) shows a pole described by a relativistic dispersion with a spectral weight proportional to its momentum. Such a collective mode can be found based on either the Kubo formula or the Boltzmann transport theory  outside the regime of the particle-hole continuum \cite{Boltzmann_LFL}. On the other hand, if the topological structure is taken into account, the nature of the zero sound mode would be modified.

In order to investigate the interplay between the topological structure of the Berry curvature and the chiral anomaly and electron correlations, we should reconcile the Boltzmann transport theory of a non-interacting Weyl metal phase \cite{CME3,CME7,Boltzmann_Chiral_Anomaly1,Boltzmann_Chiral_Anomaly2,Boltzmann_Chiral_Anomaly3,Boltzmann_Chiral_Anomaly4,Boltzmann_Chiral_Anomaly5,Boltzmann_Chiral_Anomaly6} with the Landau's Fermi-liquid theory \cite{Boltzmann_LFL}. In this study we propose a topological Fermi-liquid theory for interacting Weyl metals with time reversal symmetry breaking. First, we review the Landau's Fermi-liquid theory rather sincerely since we follow exactly the same procedure to construct a topological Fermi-liquid theory. We consider an effective free-energy functional for thermodynamics in terms of the distribution function as an order parameter and construct an effective Boltzmann transport theory for dynamic correlation functions, where only forward scattering has been introduced for spinless fermions. Reviewing Shankar's renormalization group analysis \cite{Shankar_Review}, we justify this Landau's Fermi-liquid theory to keep the role of the forward scattering only. Second, we propose a topological Fermi-liquid theory. Based on a modified band structure due to the Berry curvature, we construct an effective free-energy functional for thermodynamics in interacting Weyl metals, where only forward scattering has been taken into account. Then, we propose a topologically modified Boltzmann transport theory for dynamic correlation functions, where not only the Berry curvature and the chiral anomaly but also the forward scattering term has been incorporated. Third, we justify this topological Fermi-liquid theory. We derive a topological Fermi-gas theory in the second-quantization representation, reproducing a topologically modified Drude model given by the first-quantization approach \cite{CME4,CME6,CME7}. In order to justify this topological Fermi-gas theory, we derive a topologically modified Boltzmann transport theory for non-interacting Weyl metals, reproducing the previous studies \cite{CME3,CME4,CME6,CME7,Boltzmann_Chiral_Anomaly1,Boltzmann_Chiral_Anomaly2,Boltzmann_Chiral_Anomaly3, Boltzmann_Chiral_Anomaly4,Boltzmann_Chiral_Anomaly5,Boltzmann_Chiral_Anomaly6}. We also derive the negative longitudinal magnetoresistivity based on the Kubo formula. As far as we know, this is the first Kubo-formula based calculation for this anomalous transport coefficient given by the chiral anomaly. Finally, we perform the renormalization group analysis, following the strategy of the Landau's Fermi-liquid theory. Indeed, we find that forward scattering is the only marginal contribution above possible superconducting transition temperatures, where the Bardeen-Cooper-Schrieffer (BCS) instability \cite{BCS_Theory} is marginally relevant as the Landau's Fermi-liquid theory. In the discussion section we show how the three dimensional character differs from the two dimensional one, deriving effective field theories in two-dimensional graphene-type band structures.

Recently, we are aware of a study on ``Berry Fermi-liquid theory", where the Landau's Fermi-liquid theory has been generalized to incorporate the Berry curvature \cite{Berry_Fermi_Liquid_Theory}. This paper justifies their Berry Fermi-liquid theory based on the diagrammatic analysis \cite{Negele_Orland_Textbook} instead of the renormalization group analysis in the action level.

\section{Review on Landau's Fermi-liquid theory}

We review Landau's Fermi-liquid theory \cite{Boltzmann_LFL} for readers who are not familiar to such a theoretical framework. Readers familiar to the kinetic theory for Landau's Fermi liquids may skip this section.

\subsection{Effective field theory for thermodynamics}

We start from a partition function for interacting spinless fermions, given by
\bqa && Z = \int D c_{\bm{p}} \exp\Big[ - \int_{0}^{\beta} d \tau \Big\{ \sum_{\bm{p}} c_{\bm{p}}^{\dagger} (\partial_{\tau} - \mu + \epsilon_{\bm{p}}) c_{\bm{p}} \nn && + \sum_{\bm{q}} V_{\bm{q}} \sum_{\bm{p}} \sum_{\bm{p}'} c^{\dagger}_{\bm{p}+\bm{q}} c^{\dagger}_{\bm{p}'-\bm{q}} c_{\bm{p}'} c_{\bm{p}} \Big\} \Big] . \label{Partition_Function_Spinless_Fermions_LFL} \eqa
Here, $c_{\bm{p}}$ is a spinless fermion field with momentum $\bm{p}$, $\mu$ is a chemical potential, and $\epsilon_{\bm{p}}$ is an energy dispersion. $V_{\bm{q}}$ is an effective interaction potential with momentum transfer $\bm{q}$, assumed to be short-ranged. It is straightforward to extend this partition function to the case of spinful fermions.

Landau's Fermi-liquid theory states that this partition function should be written as follows
\bqa && Z = \exp\Big( - \beta F_{FL}[\delta n(\bm{p})] \Big) \eqa
in the low-energy limit. $F_{FL}[\delta n(\bm{p})]$ is the Landau's Fermi-liquid free-energy functional in terms of an order parameter field, here, a density-fluctuation field, defined by
\bqa && \delta n(\bm{p}) = n(\bm{p}) - n_{eq}(\bm{p}) . \eqa
\bqa && n(\bm{p}) = \langle c^{\dagger}_{\bm{p}} c_{\bm{p}} \rangle \eqa
is a distribution function, dressed by forward scattering and determined self-consistently (below), and
\bqa && n_{eq}(\bm{p}) = \frac{1}{\exp\Big\{ \beta \Big( \varepsilon_{qp}(\bm{p}) - \mu \Big) \Big\} + 1} \eqa
is an equilibrium Fermi-Dirac distribution function, where $\varepsilon_{qp}(\bm{p})$ is a quasi-particle energy dispersion with Hartree-Fock self-energy corrections. The Landau's Fermi-liquid free-energy functional is given by
\bqa && F_{FL}[\delta n(\bm{p})] = E_{FL}[\delta n(\bm{p})] - T S[n(\bm{p})] , \eqa
where the energy functional is
\bqa E_{FL}[\delta n(\bm{p})] &=& \sum_{\bm{p}} \Big( \varepsilon_{qp}(\bm{p}) - \mu \Big) \delta n(\bm{p}) \nn &+& \sum_{\bm{p}} \sum_{\bm{p}'} F_{\bm{p} \bm{p}'} \delta n(\bm{p}) \delta n(\bm{p}') \label{Energy_Functional_LFL} \eqa
and the entropy is
\bqa S[n(\bm{p})] &=& k_{B} \sum_{\bm{p}} \Big\{ n(\bm{p}) \ln n(\bm{p}) \nn &+& [1 - n(\bm{p})] \ln [1 - n(\bm{p})] \Big\} . \eqa
Here, $F_{\bm{p} \bm{p}'}$ is the Landau's interaction parameter, resulting from the effective interaction potential in Eq. (\ref{Partition_Function_Spinless_Fermions_LFL}) through renormalization. $k_{B}$ is the Boltzmann constant.

An essential point of this effective energy functional lies in the emergence of local $U(1)$ symmetry in momentum space. In other words, effective interactions between spinless fermions in Eq. (\ref{Partition_Function_Spinless_Fermions_LFL}) become simplified or renormalized as forward scattering in Eq. (\ref{Energy_Functional_LFL}), where only $\bm{q} = 0$ transfer momentum is allowed in the low-energy limit of Eq. (\ref{Partition_Function_Spinless_Fermions_LFL}) and the resulting effective interaction potential is written in terms of density fluctuations. As a result, the density at each momentum is a conserved quantity in the low-energy limit. This density-fluctuation field is determined, minimizing the free-energy functional with respect to the order-parameter field,
\bqa &&  \frac{\partial F_{FL}[\delta n(\bm{p})]}{\partial \delta n(\bm{p})} = 0 . \eqa
More explicitly, the self-consistent equation reads
\bqa && n(\bm{p}) = \frac{1}{\exp\Big\{ \beta \Big( \varepsilon_{qp}(\bm{p}) - \mu + \sum_{\bm{p}'} F_{\bm{p} \bm{p}'} \delta n(\bm{p}') \Big) \Big\} + 1} . \nn \eqa
Since the free-energy functional is given, thermodynamics of the Landau's Fermi-liquid state is determined completely \cite{Boltzmann_LFL}.

\subsection{Boltzmann transport theory for dynamic correlation functions}

\subsubsection{Boltzmann transport theory in the presence of Fermi-liquid interactions}

Dynamic correlation functions can be determined by Boltzmann transport theory in the presence of forward scattering, given by
\bqa && \partial_{t} n(\bm{p};\bm{r},t) + \bm{\dot{r}} \cdot \bm{\nabla}_{\bm{r}} n(\bm{p};\bm{r},t) + \bm{\dot{p}} \cdot \bm{\nabla}_{\bm{p}} n(\bm{p};\bm{r},t) \nn && = - \frac{n(\bm{p};\bm{r},t) - n_{eq}(\bm{p})}{\tau} . \eqa
$n(\bm{p};\bm{r},t)$ is a distribution function away from equilibrium, where $\bm{p}$ is a relative momentum, the Fourier-transformed coordinate of a relative distance between a particle-hole pair, and $\bm{r}$ ($t$) is the center-of-mass coordinate (time) of the pair \cite{Boltzmann_LFL}. $\tau$ is a relaxation time, resulting from disorder scattering.

The effective group velocity is given by
\bqa \bm{\dot{r}} &=& \bm{\nabla}_{\bm{p}} \Big( \varepsilon_{qp}(\bm{p}) - \mu + \sum_{\bm{p}'} F_{\bm{p} \bm{p}'} \delta n(\bm{p}';\bm{r},t) \Big) \nn &\approx& \bm{\nabla}_{\bm{p}} \varepsilon_{qp}(\bm{p}) \label{Drude_Model_Velocity_LFL} \eqa
in the presence of forward scattering. The effective Newton's equation is modified as follows
\bqa \bm{\dot{p}} &=& - \bm{\nabla}_{\bm{r}} \Big( \varepsilon_{qp}(\bm{p}) - \mu + \sum_{\bm{p}'} F_{\bm{p} \bm{p}'} \delta n(\bm{p}';\bm{r},t) \Big) \nn &&+ e \Big( \bm{E} + \frac{1}{c} \bm{\dot{r}} \times \bm{B} \Big) \nn &\approx& - \sum_{\bm{p}'} F_{\bm{p} \bm{p}'} \bm{\nabla}_{\bm{r}} \delta n(\bm{p}';\bm{r},t) + e \bm{E} + \frac{e}{c} [\bm{\nabla}_{\bm{p}} \varepsilon_{qp}(\bm{p})] \times \bm{B} , \nn \label{Drude_Model_Force_LFL} \eqa
where the forward-scattering term gives rise to an effective force in addition to the Lorentz force. The effective forward interaction reshuffles the density distribution in the momentum space. As a result, the Boltzmann transport equation for the Landau's Fermi-liquid state is given by
\bqa && \partial_{t} n(\bm{p};\bm{r},t) + [\bm{\nabla}_{\bm{p}} \varepsilon_{qp}(\bm{p})] \cdot \bm{\nabla}_{\bm{r}} n(\bm{p};\bm{r},t) \nn && + \Big( - \partial_{\varepsilon_{qp}} n_{eq}[\varepsilon_{qp}(\bm{p})] \Big) \sum_{\bm{p}'} F_{\bm{p} \bm{p}'} [\bm{\nabla}_{\bm{p}} \varepsilon_{qp}(\bm{p}) ] \cdot \bm{\nabla}_{\bm{r}} \delta n(\bm{p}';\bm{r},t) \nn && + \Big( e \bm{E} + \frac{e}{c} [\bm{\nabla}_{\bm{p}} \varepsilon_{qp}(\bm{p})] \times \bm{B} \Big) \cdot \bm{\nabla}_{\bm{p}} n(\bm{p};\bm{r},t) \nn && = - \frac{n(\bm{p};\bm{r},t) - n_{eq}(\bm{p})}{\tau} , \label{Boltzmann_LFL} \eqa
where $\Big( - \partial_{\varepsilon_{qp}} n_{eq}[\varepsilon_{qp}(\bm{p})] \Big) \sum_{\bm{p}'} F_{\bm{p} \bm{p}'} [\bm{\nabla}_{\bm{p}} \varepsilon_{qp}(\bm{p}) ] \cdot \bm{\nabla}_{\bm{r}} \delta n(\bm{p}';\bm{r},t)$ is an additional contribution beyond the Boltzmann transport theory of a Fermi-gas phase, resulting from reshuffling of a density distribution given by forward scattering.

\subsubsection{Drude model in the absence of interactions}

The Drude model of Eqs. (\ref{Drude_Model_Velocity_LFL}) and (\ref{Drude_Model_Force_LFL}) look trivial. However, this part should be generalized to encode the topological structure of Berry curvature and chiral anomaly in noninteracting Weyl metals. In this respect we would like to show the derivation of the Drude model for our later purpose. An electron on a Fermi surface is described by the following effective action
\bqa && S_{eff} = \int_{t_{i}}^{t_{f}}\!d t \ \Big( \bm{p} \cdot \bm{\dot{r}} + \frac{e}{c} \bm{A} \cdot \bm{\dot{r}} - e \Phi - \varepsilon_{qp}(\bm{p}) \Big) , \eqa
where $\bm{A}$ and $\Phi$ are electromagnetic vector and scalar potentials, respectively. It is straightforward to read the corresponding Hamiltonian 
\bqa && H_{eff} = - \frac{e}{c} \bm{A} \cdot \bm{\dot{r}} + e \Phi + \varepsilon_{qp}(\bm{p}) . \eqa
Hamiltonian equations of motion give rise to the Drude model, given by
\bqa && \bm{\dot{r}} = \frac{\partial H_{eff}}{\partial \bm{p}} = \bm{\nabla}_{\bm{p}} \varepsilon_{qp}(\bm{p}) \eqa
and
\bqa && \bm{\dot{p}} = - \frac{\partial H_{eff}}{\partial \bm{r}} = e \bm{E} + \frac{e}{c} \bm{\dot{r}} \times \bm{B} . \eqa

\subsubsection{Current conservation law in the presence of Fermi-liquid interactions}

One can derive the current conservation law
\bqa &&
\partial_{t} \rho(\bm{r},t) + \bm{\nabla}_{\bm{r}} \cdot \bm{j}(\bm{r},t) = 0
\eqa
from the Boltzmann transport theory Eq. (\ref{Boltzmann_LFL}) for the Landau's Fermi-liquid state. The density is
\bqa && \rho(\bm{r},t) = \int\! \frac{d^{3} \bm{p}}{(2\pi)^{3}} \ n(\bm{p};\bm{r},t) \eqa
and the electrical current is
\bqa && \bm{j}(\bm{r},t) = \int\! \frac{d^{3} \bm{p}}{(2\pi)^{3}} \ [\bm{\nabla}_{\bm{p}} \varepsilon_{qp}(\bm{p})] \Big\{ n(\bm{p};\bm{r},t) \nn && + \Big( - \partial_{\varepsilon_{qp}} n_{eq}[\varepsilon_{qp}(\bm{p})] \Big) \sum_{\bm{p}'} F_{\bm{p} \bm{p}'} \delta n(\bm{p}';\bm{r},t) \Big\} . \eqa
It is important to notice that the conserved current is modified by the forward scattering, which results from ``back flow" \cite{Boltzmann_LFL}.

\subsubsection{Density-density correlation function in the presence of Fermi-liquid interactions}

We review how to obtain dynamic correlation functions based on Boltzmann transport theory for readers who are not familiar to this framework. For simplicity, we focus on a density response, defined by
\bqa && \langle \delta \rho(\bm{q},\nu) \rangle = \chi(\bm{q},\nu) \delta \phi(\bm{q},\nu) , \eqa
where $\langle \delta \rho(\bm{q},\nu) \rangle$ is an ensemble average of density fluctuations, driven by external potential fluctuations $\delta \phi(\bm{q},\nu)$ with momentum $\bm{q}$ and frequency $\nu$. The susceptibility $\chi(\bm{q},\nu)$ is given by the density-density correlation function at equilibrium.

The left-hand-side is determined by the distribution function as follows
\bqa && \langle \delta \rho(\bm{q},\nu) \rangle \equiv \sum_{\bm{p}} \delta n(\bm{p};\bm{q},\nu) . \eqa
As a result, the compressibility is given by
\bqa && \chi(\bm{q},\nu) = \frac{\sum_{\bm{p}} \delta n(\bm{p};\bm{q},\nu)}{\delta \phi(\bm{q},\nu)} . \eqa

Turning on an external electric field represented by electric potential and performing the Fourier transformation for the center-of-mass coordinate and time, the Boltzmann equation reads
\bqa && \Big( - i \nu + \frac{1}{\tau} + [\bm{\nabla}_{\bm{p}} \varepsilon_{qp}(\bm{p})] \cdot (i \bm{q}) \Big) \delta n(\bm{p};\bm{q},\nu) \nn && + \Big( - \partial_{\varepsilon_{qp}} n_{eq}[\varepsilon_{qp}(\bm{p})] \Big) \sum_{\bm{p}'} F_{\bm{p} \bm{p}'} [\bm{\nabla}_{\bm{p}} \varepsilon_{qp}(\bm{p}) ] \cdot (i \bm{q}) \delta n(\bm{p}';\bm{q},\nu) \nn &&  = - e \Big( - \partial_{\varepsilon_{qp}} n_{eq}[\varepsilon_{qp}(\bm{p})] \Big) (i \bm{q}) \cdot [\bm{\nabla}_{\bm{p}} \varepsilon_{qp}(\bm{p})] \delta \phi(\bm{q},\nu) . \eqa
It is straightforward to solve this equation in a Fermi-gas state, where interactions are neglected. In this case the dynamic susceptibility is
\bqa && \chi_{FG}(\bm{q},\nu) \nn&&= \sum_{\bm{p}} \frac{e \Big( - \partial_{\varepsilon_{qp}} n_{eq}[\varepsilon_{qp}(\bm{p})] \Big) [\bm{\nabla}_{\bm{p}} \varepsilon_{qp}(\bm{p})] \cdot (i \bm{q})}{ i \nu - \frac{1}{\tau} - [\bm{\nabla}_{\bm{p}} \varepsilon_{qp}(\bm{p})] \cdot (i \bm{q})} . \label{Susceptibility_FG} \eqa
Although the presence of the forward scattering does not allow this calculation to be trivial, it turns out that the resulting susceptibility is given by an RPA expression \cite{Boltzmann_LFL}. A ``bare" polarization bubble corresponds to Eq. (\ref{Susceptibility_FG}), and the forward scattering strength $F_{\bm{p} \bm{p}'}$ gives rise to the geometric sum of the bare bubble. Here, we do not consider this issue more.

\subsection{Renormalization group analysis}

The effective field theory Eq. (\ref{Energy_Functional_LFL}) can be justified by the renormalization group analysis: Only the forward scattering channel turns out to be marginal for generic Fermi surfaces above the superconducting transition temperature while other interactions are irrelevant in the renormalization group sense \cite{Shankar_Review}.

Taking into account the linearized band dispersion near a Fermi surface
\bqa && \epsilon_{\bm{p}} - \mu \approx \bm{v}_{F} \cdot (\bm{p} - \bm{p}_{F}) \equiv v_{F} p , \eqa
where $\bm{v}_{F}$ is a Fermi velocity, we rewrite the partition function of Eq. (\ref{Partition_Function_Spinless_Fermions_LFL}) as follows
\begin{widetext}
\bea && Z = \int D c_{\bm{p}} \exp\Bigg[ - \int_{0}^{\beta}\! d \tau \ \bigg\{ N_{F} \int \frac{d \Omega_{d}}{S_{d}} \int_{- \Lambda}^{\Lambda} d p \ c_{\bm{p}}^{\dagger} (\partial_{\tau} + v_{F} p) c_{\bm{p}} \nn &&\hspace{120pt} + N_{F}^{4} \int\! \frac{d \Omega_{d}}{S_{d}} \int_{- \Lambda}^{\Lambda}\! d p \int\! \frac{d \Omega_{d}'}{S_{d}} \int_{- \Lambda}^{\Lambda}\! d p' \int\! \frac{d \omega_{d}}{S_{d}} \int_{- \Lambda}^{\Lambda} \! d k \int\! \frac{d \omega_{d}'}{S_{d}} \int_{- \Lambda}^{\Lambda}\! d k' \nn &&\hspace{250pt} \delta^{(d)}(\bm{p} + \bm{p}' - \bm{k} - \bm{k}') V(\bm{k},\bm{k}';\bm{p}',\bm{p}) c^{\dagger}_{\bm{k}} c^{\dagger}_{\bm{k}'} c_{\bm{p}'} c_{\bm{p}} \bigg\} \Bigg] . \eea
\end{widetext} 
Here, the integral expression $\sum_{\bm{p}} = \int \frac{d^{d} \bm{p}}{(2\pi)^{d}}$ in Eq. (\ref{Partition_Function_Spinless_Fermions_LFL}) is replaced with $N_{F} \int \frac{d \Omega_{d}}{S_{d}} \int_{- \Lambda}^{\Lambda} d p$ in the above, where the direction denoted by $\bm{\nabla}_{\bm{p}} \epsilon_{\bm{p}} / |\bm{\nabla}_{\bm{p}} \epsilon_{\bm{p}}|$ is identified with the ``radial" direction. $N_{F}$ is the density of states at the Fermi energy, assumed to be a constant value for a generic Fermi surface. $S_{d}$ is the $d-$dimensional solid angle, and $\int d \Omega_{d}$ is an angular integral. $\Lambda$ is a momentum cutoff. For interactions, the following normalization condition has been used.
\bqa && N_{F} \int \frac{d \Omega_{d}}{S_{d}} \int_{- \Lambda}^{\Lambda} d p \  \delta^{(d)}(\bm{p}) = 1 . \eqa

First of all, we should solve the $\delta-$function constraint in the interaction term, given by the momentum conservation law
\bqa && \bm{p} + \bm{p}' = \bm{k} + \bm{k}' , \eqa
where all momenta are near the Fermi surface. For simplicity, we consider $d = 2$. It turns out that only three types of scattering events are allowed \cite{Shankar_Review}: The first is forward scattering, described by
\bqa && \bm{p} = \bm{k}, ~~~~~ \bm{p}' = \bm{k}' . \eqa
The second is backward (exchange) scattering, given by
\bqa && \bm{p} = \bm{k}', ~~~~~ \bm{p}' = \bm{k} , \eqa
which is identical to forward scattering in the case of spinless fermions. The last is BCS scattering (pairing). The solution is
\bqa && \bm{p} = - \bm{p}', ~~~~~ \bm{k} = - \bm{k}' . \eqa

Considering these solutions into the partition function, we reach the following expression for the renormalization group analysis
\begin{widetext}
\bqa && W = \int D c_{\bm{p}} \exp\Bigg[ - \bigg\{ N_{F} \int \frac{d \Omega_{d}}{S_{d}} \int_{-\infty}^{\infty} \frac{d \omega}{2\pi} \int_{- \Lambda}^{\Lambda} d p \ c_{\bm{p}}^{\dagger} (- i \omega + v_{F} p) c_{\bm{p}} \nn && + N_{F}^{3} \int \frac{d \Omega_{d}}{S_{d}} \int \frac{d \Omega_{d}'}{S_{d}} \int \frac{d \omega_{d}}{S_{d}} \int_{-\infty}^{\infty} \frac{d \omega}{2\pi} \int_{-\infty}^{\infty} \frac{d \omega'}{2\pi} \int_{-\infty}^{\infty} \frac{d \Omega}{2\pi} \int_{- \Lambda}^{\Lambda} d p \int_{- \Lambda}^{\Lambda} d p' \int_{- \Lambda}^{\Lambda} d q \  F(\bm{q}) c^{\dagger}_{\bm{p}+\bm{q}} c^{\dagger}_{\bm{p}'-\bm{q}} c_{\bm{p}'} c_{\bm{p}} \nn && + N_{F}^{3} \int \frac{d \Omega_{d}}{S_{d}} \int \frac{d \Omega_{d}'}{S_{d}} \int \frac{d \omega_{d}}{S_{d}} \int_{-\infty}^{\infty} \frac{d \omega}{2\pi} \int_{-\infty}^{\infty} \frac{d \omega'}{2\pi} \int_{-\infty}^{\infty} \frac{d \Omega}{2\pi} \int_{- \Lambda}^{\Lambda} d p \int_{- \Lambda}^{\Lambda} d p' \int_{- \Lambda}^{\Lambda} d q \ V(\bm{q}) c^{\dagger}_{\bm{p}'} c^{\dagger}_{- \bm{p}' + \bm{q}} c_{- \bm{p} + \bm{q}} c_{\bm{p}} \bigg\} \Bigg] . \eqa
\end{widetext}
Here, $\int_{0}^{\beta} d \tau$ is translated into $\sum_{i \omega}$ by Fourier transformation, where $\omega$ is the Matsubara frequency for fermions, and the discrete summation is replaced with $\int_{-\infty}^{\infty} \frac{d \omega}{2\pi}$ at zero temperature. Accordingly, the symbol of the partition function $Z$ is replaced with $W$. For effective interactions, we decompose them into forward and BCS scattering channels, where $F(\bm{q})$ is the forward scattering amplitude and $V(\bm{q})$ is the BCS one.

It is straightforward to see that this effective action remains invariant under the scale transformation of
\bqa && p = \frac{p_r}{r} , ~~~~~ \omega = \frac{\omega_r}{r} , \eqa
if the fermion field transforms as follows
\bqa && c(p_{r}/r,\omega_{r}/r) = r^{\Delta_{c}} c(p_{r},\omega_{r}) \longrightarrow \Delta_{c} = \frac{3}{2} . \eqa
Here, $r$ is a scaling parameter. Both the forward and BCS scattering amplitudes do not change under this scale transformation, given by
\bqa && F(q_{r}/r) = r^{\Delta_{F}} F(q_{r}) \longrightarrow \Delta_{F} = 0 , \nn && V(q_{r}/r) = r^{\Delta_{V}} V(q_{r}) \longrightarrow \Delta_{V} = 0 . \eqa

This tree-level scaling analysis should be checked out, taking into account quantum corrections. We separate high and low energy degrees of freedoms as follows
\bqa c(p,\omega) &=& c(p_{h},\omega) \theta(|\Lambda| > |p_{h}| > |\Lambda|/r) \nn &+& c(p_{l},\omega) \theta(\Lambda/r > |p_{l}|) , \eqa
where $\theta(|\Lambda| > |p_{h}| > |\Lambda|/r) = 1$ and $\theta(\Lambda/r > |p_{l}|) = 1$ result when $|\Lambda| > |p_{h}| > |\Lambda|/r$ and $\Lambda/r > |p_{l}|$ are satisfied, respectively, and otherwise, they are zero. Then, the partition function is written by
\bqa && Z = \int D c(p_{l},\omega) D c(p_{h},\omega) \exp\Big(- S_{l}[c(p_{l},\omega)] \nn && - S_{h}[c(p_{h},\omega)] - S_{int}[c(p_{l},\omega),c(p_{h},\omega)] \Big) \eqa
in terms of these high and low energy degrees of freedom. Here,
\bqa S_{l}[c(p_{l},\omega)] &=& N_{F} \int \frac{d \Omega_{d}}{S_{d}} \int_{-\infty}^{\infty} \frac{d \omega}{2\pi} \int_{- \Lambda/r}^{\Lambda/r} d p_{l} \nn && c^{\dagger}(p_{l},\omega)  (- i \omega + v_{F} p_{l}) c(p_{l},\omega) \eqa
is an effective action for low energy fermions, and
\bqa S_{h}[c(p_{h},\omega)] &=& N_{F} \int \frac{d \Omega_{d}}{S_{d}} \int_{-\infty}^{\infty} \frac{d \omega}{2\pi} \int_{|\Lambda|/r}^{|\Lambda|} d p_{h} \nn && c^{\dagger}(p_{h},\omega)  (- i \omega + v_{F} p_{h}) c(p_{h},\omega) \eqa
is that for high energy fermions. The interaction part between high and low energy fermions is described by the following effective action
\begin{widetext}
\bqa && S_{int}[c(p_{l},\omega),c(p_{h},\omega)] \nn && = N_{F}^{3} \int \frac{d \Omega_{d}}{S_{d}} \int \frac{d \Omega_{d}'}{S_{d}} \int \frac{d \omega_{d}}{S_{d}} \int_{-\infty}^{\infty} \frac{d \omega}{2\pi} \int_{-\infty}^{\infty} \frac{d \omega'}{2\pi} \int_{-\infty}^{\infty} \frac{d \Omega}{2\pi} \int_{- \Lambda/r}^{\Lambda/r} d p_{l} \int_{|\Lambda|/r}^{|\Lambda|} d p_{h}' \int_{- \Lambda}^{\Lambda} d q \nn &&\hspace{180pt} F(q) c^{\dagger}(p_{l}+q,\omega+\Omega) c^{\dagger}(p_{h}'-q,\omega'-\Omega) c(p_{h}',\omega') c(p_{l},\omega) \nn && + N_{F}^{3} \int \frac{d \Omega_{d}}{S_{d}} \int \frac{d \Omega_{d}'}{S_{d}} \int \frac{d \omega_{d}}{S_{d}} \int_{-\infty}^{\infty} \frac{d \omega}{2\pi} \int_{-\infty}^{\infty} \frac{d \omega'}{2\pi} \int_{-\infty}^{\infty} \frac{d \Omega}{2\pi} \int_{- \Lambda/r}^{\Lambda/r} d p_{l} \int_{|\Lambda|/r}^{|\Lambda|} d p_{h}' \int_{- \Lambda}^{\Lambda} d q  \nn &&\hspace{180pt} V(q) c^{\dagger}(p_{h}',\omega') c^{\dagger}(-p_{h}'+q,-\omega'+\Omega) (-p_{l}+q,-\omega+\Omega) c(p_{l},\omega) , \eqa
\end{widetext}
where the transfer momentum $\bm{q}$ is vanishingly small.

The next step is to integrate over high energy fermion fields. Then, the effective partition function reads
\bqa && Z_{eff} = \int\! D c(p_{l},\omega) \exp\Big(- S_{l}[c(p_{l},\omega)] - \mathcal{S}^{(2)}[c(p_{l},\omega)] \Big) \nn \eqa
in terms of low energy fermion fields. The interaction term between low energy fermions can be found in the second-order cumulant expansion, given by
\bqa \mathcal{S}^{(2)}[c(p_{l},\omega)] &=& - \frac{1}{2} \bigg( \big\langle S_{int}^{2}[c(p_{l},\omega),c(p_{h},\omega)] \big\rangle_{h} \nn &-& \big\langle S_{int}[c(p_{l},\omega),c(p_{h},\omega)] \big\rangle_{h}^{2} \bigg) , \eqa
where the average with the subscript $h$ is defined by
\bqa && \big\langle \mathcal{O}[c(p_{l},\omega),c(p_{h},\omega)] \big\rangle_{h} = \frac{1}{Z} \int\! D c(p_{h},\omega) \nn && \mathcal{O}[c(p_{l},\omega),c(p_{h},\omega)] \exp\Big( - S_{h}[c(p_{h},\omega)] \Big) . \eqa

It turns out that the forward scattering channel still remains marginal
\bqa && \frac{d F(r)}{d \ln r} = 0 , \eqa
even if quantum corrections are taken into account \cite{Shankar_Review}. On the other hand, the BCS pairing channel becomes marginally relevant, described by \cite{Shankar_Review}
\bqa && \frac{d V(r)}{d \ln r} = - c N_{F} V^{2}(r) , \eqa
where $c$ is a positive numerical constant. The solution is
\bqa && V(T) = \frac{V}{1- c N_{F} V \ln(D/T)} , \eqa
where $D$ is a bandwidth, defining UV. Attractive interactions at UV enhance to be infinite at IR, implying the BCS superconducting instability at the critical temperature \cite{BCS_Theory}
\bqa && T_{c} = D \exp\Big(- \frac{1}{c N_{F} V} \Big) . \eqa
In this respect the Landau's Fermi-liquid state is a stable fixed point of ``weakly" interacting fermions above the superconducting transition temperature, described by the Landau's Fermi-liquid theory discussed before.

\section{Topological Fermi-liquid theory}

We repeat exactly what we have discussed before in the Landau's Fermi-liquid theory for interacting Weyl metals. We extend the existing topological Fermi-gas theory for noninteracting Weyl metals to a topological Fermi-liquid theory for interacting Weyl metals.

\subsection{Effective field theory for a Weyl metallic state}

First, we need to discuss an effective field theory as our starting point, corresponding to Eq. (\ref{Partition_Function_Spinless_Fermions_LFL}). We start from an effective Dirac theory with an inhomogeneous topological-in-origin $\theta-$term
\begin{widetext}
\bqa && Z = \int\! D \psi_{\alpha a} \exp\Bigg[- \int_{0}^{\beta}\! d \tau \int\! d^{3} \bm{r} \ \bigg\{ \psi_{\alpha a}^{\dagger} \Big( (\partial_{\tau} - \mu) \bm{I}_{\alpha\beta} \otimes \bm{I}_{ab} - i v_{D} \Big(\bm{\partial}_{\bm{r}} - i\frac{e}{c} \bm{A} \Big) \cdot \bm{\sigma}_{\alpha\beta} \otimes \bm{\tau}_{ab}^{z} + m \bm{I}_{\alpha\beta} \otimes \bm{\tau}_{ab}^{x} \Big) \psi_{\beta b} \nn && + \frac{1}{8 \pi} (\bm{E}^{2} + \bm{B}^{2}) + \frac{\theta(\bm{r})}{2\pi} \frac{\alpha}{2\pi} \bm{E} \cdot \bm{B} + V(\phi,\varphi) \Big[\psi_{\alpha a}^{\dagger} (\sin \phi \bm{I}_{\alpha\beta} + \cos \phi {\bm n}_\phi \cdot \bm{\sigma}_{\alpha\beta}) \otimes (\sin \varphi \bm{I}_{ab} + \cos \varphi {\bm n}_\varphi \cdot \bm{\tau}_{ab}) \psi_{\beta b}\Big]^{2} \bigg\} \Bigg] . \nn \label{TFL_Dirac_Theory} \eqa
\end{widetext}
Here, $\psi_{\alpha a}$ is a four-component Dirac spinor with spin $\alpha$ and orbital $a$. $\bm{\sigma}_{\alpha\beta}$ and $\bm{\tau}_{ab}$ are two-by-two Pauli matrices, acting on spin and orbital spaces, respectively. $v_{D}$ is a velocity, $m$ is a mass parameter, and $\mu$ is a chemical potential. $\bm{A}$ is an electromagnetic vector potential, regarded to be externally applied.
\bqa && \bm{E} = - \frac{1}{c} \partial_{\tau} \bm{A}, ~~~~~ \bm{B} = \bm{\nabla} \times \bm{A} \eqa
are externally applied electric field and magnetic field, respectively. $\alpha$ is a fine structure constant, and $\theta(\bm{r})$ is an axion angle which is determined by the strength of an external magnetic field as shown below. The last term describes effective interactions between Dirac fermions, generally expressed.

One can represent this effective theory in terms of four-by-four Dirac gamma matrices, given by
\bqa && \gamma^{0} = \bm{I}_{\alpha\beta} \otimes \bm{\tau}_{ab}^{x} , ~~~~~ \gamma^{k} = - i \bm{\sigma}_{\alpha\beta}^{k} \otimes \bm{\tau}_{ab}^{y} . \eqa
Then, the partition function reads
\begin{widetext}
\bqa && Z = \int\! D \psi \ \exp\Bigg[- \int_{0}^{\beta}\! d \tau \int\! d^{3} \bm{r} \ \bigg\{ \bar{\psi} \Big( i \gamma^{0} (\partial_{\tau} - \mu) - i v_{D} \bm{\gamma} \cdot \Big(\bm{\partial}_{\bm{r}} - i \frac{e}{c} \bm{A}\Big) + m \Big) \psi + \frac{1}{8 \pi} (\bm{E}^{2} + \bm{B}^{2}) + \frac{\theta(\bm{r})}{2\pi} \frac{\alpha}{2\pi} \bm{E} \cdot \bm{B} \nn &&\hspace{130pt} + \lambda_{s} (\bar{\psi} \psi)^{2} + \lambda_{v} (\bar{\psi} \gamma^{\mu} \psi)^{2} + \lambda_{as} (\bar{\psi} \gamma^{\mu\nu} \psi)^{2} + \lambda_{pv} (\bar{\psi} \gamma^{\mu} \gamma^{5} \psi)^{2} + \lambda_{ps} (\bar{\psi} \gamma^{5} \psi)^{2} \bigg\} \Bigg] , \label{TFL_Dirac_Theory_Interactions} \eqa
\end{widetext}
where
\bqa && \bar{\psi} = \psi^{\dagger} \gamma^{0} . \eqa
Effective interactions between Dirac fermions are decomposed systematically into scalar $(1)$ $\oplus$ vector $(4)$ $\oplus$ antisymmetric tensor $(6)$ $\oplus$ pseudovector $(4)$ $\oplus$ pseudoscalar $(1)$, denoted by the strengths of effective interactions $\lambda_{s}$, $\lambda_{v}$, $\lambda_{as}$, $\lambda_{pv}$, and $\lambda_{ps}$, respectively. Recall $1 + 4 + 6 + 4 + 1 = 16$, implying that any four-by-four matrices are described by this basis. The antisymmetric tensor is
\bqa && \gamma^{\mu\nu} = \frac{1}{2} [\gamma^{\mu},\gamma^{\nu}] , \eqa
and the chiral matrix is
\bqa && \gamma^{5} = i \gamma^{0} \gamma^{1} \gamma^{2} \gamma^{3} = \bm{I}_{\alpha\beta} \otimes \bm{\tau}_{ab}^{z} . \eqa

In order to determine the angle parameter, we recall the chiral anomaly equation
\bqa && \partial_{\mu} (\bar{\psi} \gamma^{\mu} \gamma^{5} \psi) = \frac{\alpha}{4 \pi^{2}} \bm{E} \cdot \bm{B} . \eqa
This equation states that the classically conserved chiral current given by the subtraction of the left-handed chiral current from the right-handed chiral current is not conserved any more in the quantum level, described by the right-hand side \cite{Peskin_Schroeder}. Replacing the topological-in-origin $\bm{E} \cdot \bm{B}$ term with the chiral current based on this anomaly equation and performing the integration by parts, we rewrite the effective action as follows
\begin{widetext}
\bqa && \mathcal{S}_{eff} = \int_{0}^{\beta}\! d \tau \int\! d^{3} \bm{r} \ \bigg\{ \bar{\psi} \Big( i \gamma^{0} (\partial_{\tau} - \mu) - i v_{D} \bm{\gamma} \cdot \Big(\bm{\partial}_{\bm{r}} - i\frac{e}{c} \bm{A} - i \gamma^{5} \bm{c}\Big) + m \Big) \psi + \frac{1}{8 \pi} (\bm{E}^{2} + \bm{B}^{2}) \nn &&\hspace{100pt} + \lambda_{s} (\bar{\psi} \psi)^{2} + \lambda_{v} (\bar{\psi} \gamma^{\mu} \psi)^{2} + \lambda_{as} (\bar{\psi} \gamma^{\mu\nu} \psi)^{2} + \lambda_{pv} (\bar{\psi} \gamma^{\mu} \gamma^{5} \psi)^{2} + \lambda_{ps} (\bar{\psi} \gamma^{5} \psi)^{2} \bigg\} . \label{eq:TFL_action}\eqa
\end{widetext}
Here, $\bm{c}$ is a chiral gauge field, given by
\bqa && \bm{c} = \bm{\nabla}_{\bm{r}} \theta(\bm{r}) . \eqa
It is essential to observe that the chiral gauge-field term is nothing but the Zeeman term, given by
\bqa && \mathcal{S}_{Z} = - \int_{0}^{\beta}\! d \tau \int d^{3} \bm{r} \ v_{D} g \psi_{\alpha a}^{\dagger} [\bm{B} \cdot (\bm{\sigma}_{\alpha\beta} \otimes \bm{I}_{ab})] \psi_{\beta b} , \nn \eqa
if this term is expressed in terms of Dirac gamma matrices. Here, $g$ is the Land\'e $g-$factor. The chiral gauge field is identified with the external magnetic field
\bqa && \bm{c} = g \bm{B}. \eqa
As a result, the axion angle is
\bqa && \theta(\bm{r}) = g \bm{B} \cdot (\bm{r} + \bm{R}) , \eqa
where $\bm{R}$ describes the freedom of a reference point.

\subsection{Topological Fermi-liquid theory for thermodynamics}

Solving Eq. (\ref{TFL_Dirac_Theory}), we find
\bqa && Z = \exp\Big( - \beta \mathcal{F}_{TFL}[\delta n_{\chi}(\bm{p})] \Big) , \eqa
where
\bqa && \mathcal{F}_{TFL}[\delta n_{\chi}(\bm{p})] = \mathcal{E}_{TFL}[\delta n_{\chi}(\bm{p})] - T \mathcal{S}_{TFL}[\delta n_{\chi}(\bm{p})] \nn \eqa
is an effective topological Fermi-liquid free-energy functional in terms of a density order parameter of spinless fermions on a pair of chiral Fermi surfaces. Here, spinless fermions appear from spin-momentum locking, which will be derived below.
\bqa && \delta n_{\chi}(\bm{p}) = n_{\chi}(\bm{p}) - n_{\chi}^{eq}(\bm{p}) \eqa
is a density-fluctuation field of spinless fermions on the chiral Fermi surface $\chi$ with momentum $\bm{p}$, where
\bqa && n_{\chi}^{eq}(\bm{p}) = \frac{1}{e^{\beta \varepsilon_{\bm{p}}^{\chi}} + 1} \eqa
is an equilibrium distribution function.
\bqa && \varepsilon_{\bm{p}}^{\chi} = \Bigl( \bm{v}_{F}^{\chi} + \frac{e}{c} (\bm{\mathcal{B}}_{F}^{\chi} \cdot \bm{v}_{F}^{\chi}) \bm{B} \Bigr) \cdot \bm{p} + \frac{e}{c} (\bm{v}_{F}^{\chi} \times \bm{\mathcal{A}}_{F}^{\chi}) \cdot \bm{B} \nn \eqa
is the energy dispersion relation for spinless fermions near a pair of chiral Fermi surfaces, modified from contributions of high energy electron fields near the pair of Weyl points. The group velocity $\bm{v}_{F}^{\chi}$ is renormalized by such high energy electrons as $\bm{v}_{F}^{\chi} + \frac{e}{c} (\bm{\mathcal{B}}_{F}^{\chi} \cdot \bm{v}_{F}^{\chi}) \bm{B}$, where $\bm{\mathcal{B}}_{F}^{\chi}$ is the Berry magnetic field at the Fermi surface of $\chi$. This modification may be interpreted as a coupling term between an emergent magnetic dipole moment and an external magnetic field, where the magnetic dipole moment originates from the existence of the Berry curvature \cite{CME6,Berry_Fermi_Liquid_Theory,Magnetic_Moment_Lorentz_Symmetry1,Magnetic_Moment_Lorentz_Symmetry2}. The last effective potential with the Berry gauge field $\bm{\mathcal{A}}_{F}^{\chi}$ describes the contribution of electric polarization. This energy dispersion will be derived in the path-integral representation below.

The energy functional for a topological Fermi-liquid state is given by
\bqa
&&\mathcal{E}_{TFL}[\delta n_{\chi}(\bm{p})] = \sum_{\bm{p}} \sum_{\chi = \pm} \varepsilon_{\bm{p}}^{\chi} \delta n_{\chi}(\bm{p}) \nn &+& \frac{1}{2} \sum_{\bm{p} \not= \bm{p}'} \sum_{\chi, \chi' = \pm} F_{\chi\chi'}(\bm{p},\bm{p}') \delta n_{\chi}(\bm{p}) \delta n_{\chi'}(\bm{p}') , \label{Energy_Functional_TFL}
\eqa
where $F_{\chi\chi'}(\bm{p},\bm{p}')$ is the Landau interaction parameter, describing the strength of forward scattering. Performing the renormalization group analysis, we will show that only the forward scattering amplitude is marginal above the critical temperature of possible superconducting phases while other interactions are irrelevant, essentially identical to the case of the Landau's Fermi-liquid theory.
%
%
\bqa \mathcal{S}_{TFL} &=& \sum_{\bm{p}} \sum_{\chi = \pm} \Big\{ n_{\chi}(\bm{p}) \ln n_{\chi}(\bm{p}) \nn &+& \Big(1 - n_{\chi}(\bm{p})\Big) \ln \Big(1 - n_{\chi}(\bm{p})\Big) \Big\},  \eqa
is the same entropy as that of the Landau's Fermi-liquid theory. 

%
%

The density order parameter is determined by the self-consistent equation of
\bqa && \frac{\delta}{\delta n_{\chi}(\bm{p};\bm{r},t)} \mathcal{F}_{TFL}[\delta n_{\chi}(\bm{p})] = 0 . \eqa
More explicitly, the self-consistent equation reads
\bqa && n_{\chi}(\bm{p}) \nn &&= f\left( \varepsilon_{\bm{p}}^{\chi} + \sum_{\bm{p}'} \sum_{\chi' = \pm} F_{\chi\chi'}(\bm{p},\bm{p}') \delta n_{\chi'}(\bm{p}') \right) \eqa
for each chiral Fermi surface, where $f(x) = 1/(e^{\beta x} +1)$ is the Fermi distribution function. As a result, we can understand the thermodynamics of a topological Fermi-liquid state. In particular, we predict that the thermodynamics in interacting Weyl metals will show unconventional magnetic-field dependence beyond the Landau's Fermi-liquid state. For example, the Wilson's ratio would be modified, compared with that of the Landau's Fermi-liquid state.

\subsection{Topological Boltzmann transport theory for dynamic correlation functions}

In order to understand dynamic correlation functions in interacting Weyl metals, it is essential to construct Boltzmann transport theory for a topological Fermi-liquid theory, generalizing either that of Landau's Fermi-liquid theory \cite{Boltzmann_LFL} with the introduction of both the Berry curvature and chiral anomaly or that of a topological Fermi-gas theory \cite{CME3,CME7,Boltzmann_Chiral_Anomaly1,Boltzmann_Chiral_Anomaly2,Boltzmann_Chiral_Anomaly3, Boltzmann_Chiral_Anomaly4,Boltzmann_Chiral_Anomaly5,Boltzmann_Chiral_Anomaly6} with the introduction of forward scattering. The Boltzmann equation is given by
\bqa &&
\partial_{t} n_{\chi}(\bm{p};\bm{r},t) + \bm{\dot{r}}_{\chi} \cdot \bm{\nabla}_{\bm{r}} n_{\chi}(\bm{p};\bm{r},t) + \bm{\dot{p}}_{\chi} \cdot \bm{\nabla}_{\bm{p}} n_{\chi}(\bm{p};\bm{r},t) \nn && = I_{coll} [\delta n_{\chi}(\bm{p};\bm{r},t)] ,
\eqa
as discussed before.
%
%
There is an additional term to modify the Drude model in the Landau's Fermi-liquid theory, encoding the topological information of both the Berry curvature and chiral anomaly. In particular, the group velocity is generalized as follows
\bqa \bm{\dot{r}}_{\chi} &=& \bm{\nabla}_{\bm{p}} \Big( \varepsilon_{\bm{p}}^{\chi} + \sum_{\bm{p}'} \sum_{\chi' = \pm} F_{\chi\chi'}(\bm{p},\bm{p}') \delta n_{\chi'}(\bm{p}';\bm{r},t) \Big) \nn &&+ \bm{\dot{p}}_{\chi} \times \boldsymbol{\mathcal{B}}_F^\chi \nn &\approx& \bm{\nabla}_{\bm{p}} \Big\{ \Bigl( \bm{v}_{F}^{\chi} + \frac{e}{c} (\bm{\mathcal{B}}_{F}^{\chi} \cdot \bm{v}_{F}^{\chi}) \bm{B} \Bigr) \cdot \bm{p} + \frac{e}{c} (\bm{v}_{F}^{\chi} \times \bm{\mathcal{A}}_{F}^{\chi}) \cdot \bm{B} \Big\} \nn &&+ \bm{\dot{p}}_{\chi} \times \boldsymbol{\mathcal{B}}_F^\chi \nn &\approx& \bm{v}_{F}^{\chi} + \frac{e}{c} (\bm{\mathcal{B}}_{F}^{\chi} \cdot \bm{v}_{F}^{\chi}) \bm{B} + \bm{\dot{p}}_{\chi} \times \boldsymbol{\mathcal{B}}_F^\chi , \eqa
where an anomalous-velocity term $\bm{\dot{p}}_{\chi} \times \boldsymbol{\mathcal{B}}_F^\chi$ appears \cite{Berry_Curvature_Review1,Berry_Curvature_Review2}, involved with the Berry curvature $\boldsymbol{\mathcal{B}}_F^\chi$. On the other hand, the force law remains unchanged as follows
\bqa \bm{\dot{p}}_{\chi} &=& - \bm{\nabla}_{\bm{r}} \Big( \varepsilon_{\bm{p}}^{\chi} + \sum_{\bm{p}'} \sum_{\chi' = \pm} F_{\chi\chi'}(\bm{p},\bm{p}') \delta n_{\chi'}(\bm{p}';\bm{r},t) \Big) \nn &&+ e \Big( \bm{E} + \frac{1}{c} \bm{\dot{r}}_{\chi} \times \bm{B} \Big) \nn &=& - \sum_{\bm{p}'} \sum_{\chi' = \pm} F_{\chi\chi'}(\bm{p},\bm{p}') \bm{\nabla}_{\bm{r}} \delta n_{\chi'}(\bm{p}';\bm{r},t) \nn &&+ e \Big( \bm{E} + \frac{1}{c} \bm{\dot{r}}_{\chi} \times \bm{B} \Big) . \eqa
We will review the derivation of these Hamilton's equations of motion below in the absence of forward scattering. Our work is to generalize this previous study in the presence of interactions, deriving these equations from Eq. (\ref{TFL_Dirac_Theory}).

It is straightforward to solve such Hamilton's equations of motion. As a result, we obtain
\bqa \bm{\dot{r}}_{\chi} &=& G_{\chi} \Big( \bm{v}_{F}^{\chi} + e {\bm E} \times \boldsymbol{\mathcal{B}}_F^\chi + \frac{e}{c} (\bm{\mathcal{B}}_{F}^{\chi} \cdot \bm{v}_{F}^{\chi}) \bm{B} \nn &-& \sum_{\bm{p}'} \sum_{\chi' = \pm} F_{\chi\chi'}(\bm{p},\bm{p}') [\bm{\nabla}_{\bm{r}} \delta n_{\chi'}(\bm{p}';\bm{r},t)] \times \boldsymbol{\mathcal{B}}_F^\chi \Big) \nn \eqa
and
\bqa \bm{\dot{p}}_{\chi} &=& G_{\chi} \Big( e \bm{E} + \frac{e}{c} \bm{v}_{F}^{\chi} \times \bm{B} + \frac{e^2}{c}( {\bm E} \cdot {\bm B} ) \boldsymbol{\mathcal{B}}_F^\chi \nn &-& \sum_{\bm{p}'} \sum_{\chi' = \pm} F_{\chi\chi'}(\bm{p},\bm{p}') \bm{\nabla}_{\bm{r}} \delta n_{\chi'}(\bm{p}';\bm{r},t) \Big) , \eqa
where
\bqa && G_{\chi} = \Big(1 + \frac{e}{c} \boldsymbol{\mathcal{B}}_F^\chi \cdot {\bm B} \Big)^{-1} \eqa
is a modification factor for the measure of a phase-space volume. Compared with the group velocity in the Landau's Fermi-liquid state, it acquires three types of corrections. $e {\bm E} \times \boldsymbol{\mathcal{B}}_F^\chi$ gives rise to an anomalous Hall effect \cite{AHE1,AHE2,AHE3,Berry_Curvature_Review1,Berry_Curvature_Review2}, regarded to be an extended version from two dimensions to three dimensions. $\frac{e}{c} (\bm{\mathcal{B}}_{F}^{\chi} \cdot \bm{v}_{F}^{\chi}) \bm{B}$ is responsible for the so called chiral magnetic effect \cite{CME1,CME2,CME3,CME4,CME5,CME6,CME7,Boltzmann_Chiral_Anomaly1,Boltzmann_Chiral_Anomaly2,Boltzmann_Chiral_Anomaly3,Boltzmann_Chiral_Anomaly4,
Boltzmann_Chiral_Anomaly5,Boltzmann_Chiral_Anomaly6}, not discussed here. These two contributions have been well known. The last term $- \sum_{\bm{p}'} \sum_{\chi' = \pm} F_{\chi\chi'}(\bm{p},\bm{p}') [\bm{\nabla}_{\bm{r}} \delta n_{\chi'}(\bm{p}';\bm{r},t)] \times \boldsymbol{\mathcal{B}}_F^\chi$ involved with the forward-scattering strength and the Berry curvature is our novel suggestion. The effect of this term will be discussed below. Compared with the force law in the Landau's Fermi-liquid state, it contains an additional term ${e^2\over c}( {\bm E} \cdot {\bm B} ) \boldsymbol{\mathcal{B}}_F^\chi$. The role of this term is well understood, responsible for the chiral anomaly in this semiclassical description \cite{CME1,CME2,CME3,CME4,CME5,CME6,CME7,Boltzmann_Chiral_Anomaly1,Boltzmann_Chiral_Anomaly2,Boltzmann_Chiral_Anomaly3,Boltzmann_Chiral_Anomaly4,Boltzmann_Chiral_Anomaly5,Boltzmann_Chiral_Anomaly6}.

The collision term is given by
\bqa I_{coll} [\delta n_{\chi}(\bm{p};\bm{r},t)] &=& - \frac{n_{\chi}(\bm{p};\bm{r},t) - n_{\chi}^{eq}(\bm{p})}{\tau_{eff}} \eqa
in the relaxation-time approximation, where $\tau_{eff}$ is an effective relaxation time to include both contributions of intra- and inter- chiral Fermi surfaces. As a result, we find
\begin{widetext}
\bqa &&
\partial_{t} n_{\chi}(\bm{p};\bm{r},t) \nn && + G_{\chi} \Big( \bm{v}_{F}^{\chi} + e {\bm E} \times \boldsymbol{\mathcal{B}}_F^\chi + \frac{e}{c} (\bm{\mathcal{B}}_{F}^{\chi} \cdot \bm{v}_{F}^{\chi}) \bm{B} \Big) \cdot \bm{\nabla}_{\bm{r}} n_{\chi}(\bm{p};\bm{r},t) + G_{\chi} \Big( e \bm{E} + \frac{e}{c} \bm{v}_{F}^{\chi} \times \bm{B} + \frac{e^2}{c}( {\bm E} \cdot {\bm B} ) \boldsymbol{\mathcal{B}}_F^\chi \Big) \cdot \bm{\nabla}_{\bm{p}} n_{\chi}(\bm{p};\bm{r},t) \nn && + [ - \partial_{\varepsilon} n_{eq}(\varepsilon) ] G_{\chi} \sum_{\bm{p}'} \sum_{\chi' = \pm} F_{\chi\chi'}(\bm{p},\bm{p}') \Big( \bm{v}_{F}^{\chi} + \frac{e}{c} (\bm{\mathcal{B}}_{F}^{\chi} \cdot \bm{v}_{F}^{\chi}) \bm{B} \Big) \cdot \bm{\nabla}_{\bm{r}} \delta n_{\chi'}(\bm{p}';\bm{r},t) = - \frac{n_{\chi}(\bm{p};\bm{r},t) - n_{\chi}^{eq}(\bm{p})}{\tau_{eff}} , \label{Boltzmann_Transport_Theory_Weyl_Metal}
\eqa
valid in the linear-response regime, where
\bqa && - G_{\chi} \Big( \sum_{\bm{p}'} \sum_{\chi' = \pm} F_{\chi\chi'}(\bm{p},\bm{p}') [\bm{\nabla}_{\bm{r}} \delta n_{\chi'}(\bm{p}';\bm{r},t)] \times \boldsymbol{\mathcal{B}}_F^\chi \Big)\cdot \bm{\nabla}_{\bm{r}} n_{\chi}(\bm{p};\bm{r},t) \nonumber \eqa
is neglected. Here, we point out an interaction-driven term
\bqa && [ - \partial_{\varepsilon} n_{eq}(\varepsilon) ] G_{\chi} \sum_{\bm{p}'} \sum_{\chi' = \pm} F_{\chi\chi'}(\bm{p},\bm{p}') \Big( \bm{v}_{F}^{\chi} + \frac{e}{c} (\bm{\mathcal{B}}_{F}^{\chi} \cdot \bm{v}_{F}^{\chi}) \bm{B} \Big) \cdot \bm{\nabla}_{\bm{r}} \delta n_{\chi'}(\bm{p}';\bm{r},t) ,
\eqa
\end{widetext}
essential to generalize the Boltzmann transport theory of the topological Fermi-gas state. Compared with the Boltzmann transport theory of the Landau's Fermi-liquid state, the Berry curvature gives rise to anisotropy for the angular dependence in the forward scattering strength. As a result, correlation functions given by RPA within the Boltzmann transport theory will show unconventional angular dependence, implying possible appearances for novel instabilities as the interaction parameter increases. For example, we predict an anisotropic dispersion relation in the zero sound mode. Response functions based on this Boltzmann transport theory are on progress.

\subsection{Review on the topological Drude model in the absence of Fermi-liquid interactions}

We review the derivation of the topologically modified Drude model \cite{CME4,CME6}. We start from an effective Weyl Hamiltonian
\bqa && H = {\bm \sigma} \cdot {\bm p} , \eqa
where $\bm{\sigma}$ is a Pauli spin matrix. Here we take $\chi=+1$ chirality for definiteness. Then, the transition amplitude is given by
\bqa
&& \langle f | e^{-iH(t_f - t_i)} | i \rangle \nn && = \int_{\bm{r}_{i}}^{\bm{r}_{f}}\! D \bm{r} \!\int\! D \bm{p} \ \exp \Big\{ i\! \int_{t_i}^{t_f}\! dt \ ( {\bm p} \cdot \dot{\bm r} - {\bm \sigma} \cdot {\bm p} ) \Big\} .
\eqa

Usually speaking, we are allowed to focus on low energy electrons near a Fermi surface without the information of high energy electrons deep inside the Fermi surface. Actually, the Drude model for a conventional Fermi surface has been derived in that way. However, the existence of the magnetic monopole and anti-monopole pair in momentum space does not allow us to take low energy electrons near a pair of chiral Fermi surfaces only. Instead, we should introduce the information of high energy electrons near the pair of Weyl points into the low energy dynamics of electrons near the pair of chiral Fermi surfaces. This UV information deep inside the Fermi surface encodes the topological structure of both the Berry curvature and the chiral anomaly into chiral electrons on the Fermi surface. This can be achieved by the integration of high energy electrons in the second-quantization representation. In the first-quantization representation we should describe the low energy dynamics in terms of the basis function that diagonalizes the Hamiltonian. In this respect we introduce the unitary matrix $U_{\bm p}$ to diagonalize the Weyl Hamiltonian,
\bqa && U_{\bm p}^\dagger {\bm \sigma} \cdot {\bm p} U_{\bm p} = |{\bm p}| \sigma^3 . \eqa
As a result, we rewrite the transition amplitude in the following way
\bqa
&& \langle f | e^{-iH(t_f - t_i)} | i \rangle = \int_{\bm{r}_{i}}^{\bm{r}_{f}} D \bm{r} \int D \bm{p} \nn && U_{{\bm p}_f}^\dagger \exp \Big\{ i\!\int_{t_i}^{t_f}\! dt \ ( {\bm p} \cdot \dot{\bm r} - |{\bm p}| \sigma^3 - \bm{\mathcal{\hat A}}_{\bm p} \cdot \dot {\bm p} ) \Big\} U_{{\bm p}_i} ,
\eqa
where
\bqa && \bm{\mathcal{\hat A}}_{\bm p} = i U_{\bm p}^\dagger {\bm \nabla}_{\bm p} U_{\bm p} \eqa
is an emergent Berry gauge field \cite{CME4}. Since we consider that the chemical potential lies much above the Weyl point, we keep only the $11-$component in the two-by-two matrix. Then, we obtain the following semiclassical effective action \bqa &&
S_{eff} = \int_{t_i}^{t_f}\! dt \ ( {\bm p} \cdot \dot{\bm r} - |{\bm p}| - \bm{\mathcal{A}}_{\bm p} \cdot \dot{\bm p} ) ,
\eqa
where
\bqa && \bm{\mathcal{A}}_{\bm p} = \Big[\bm{\mathcal{\hat A}}_{\bm p}\Big]_{11} \eqa
is the Berry gauge field.

Introducing electromagnetic fields into the above and performing the path-integral representation within the diagonal basis carefully, in particular, keeping the Lorentz symmetry, one can find an effective semiclassical action \cite{CME6}
\bqa S_{eff} &=& \int_{t_i}^{t_f}\! dt \ \Big\{ \Big( {\bm p} + \frac{e}{c} {\bm A} \Big) \cdot \dot{\bm r} -  e\varphi - \bm{\mathcal{A}}_{\bm p} \cdot \dot{\bm p} \nn &&- \Big( 1 + \frac{e}{c} \bm{\mathcal{B}}_{\bm p} \cdot {\bm B} \Big)  |{\bm p}| \Big\} .
\eqa
Here, ${\bm A}$ and $\varphi$ are electromagnetic vector and scalar potentials, respectively. $\bm{\mathcal{B}}_{\bm p}$ is the Berry magnetic field, given by
\bqa && \bm{\mathcal{B}}_{\bm p} = {\bm \nabla}_{\bm p} \times \bm{\mathcal{A}}_{\bm p} . \eqa
We recall that the group velocity is modified to $1 + \bm{\mathcal{B}}_{\bm p} \cdot {\bm B}$, as discussed before. It is straightforward to find the corresponding Hamiltonian as follows
\bqa && H_{eff} = - \frac{e}{c} {\bm A} \cdot \dot{\bm r} +  e\varphi + \bm{\mathcal{A}}_{\bm p} \cdot \dot{\bm p} + \Big( 1 + \frac{e}{c} \bm{\mathcal{B}}_{\bm p} \cdot {\bm B} \Big) |{\bm p}| . \nn \eqa
The Hamiltonian equation of motion now reads
\bqa && \dot{\bm r} = \frac{\partial H_{eff}}{\partial \bm{p}} = {\bm v}_F + \dot{\bm p} \times \bm{\mathcal{B}}_F , \nn && \dot{\bm p} = - \frac{\partial H_{eff}}{\partial \bm{r}} = e{\bm E} + \frac{e}{c} \dot{\bm r} \times {\bm B} , \eqa
where $\bm{\mathcal{B}}_F \equiv \bm{\mathcal{B}}_{\bm p} |_{\bm p = \bm p_F}$. Here, the anomalous-velocity term $\dot{\bm p} \times \bm{\mathcal{B}}_F$ appears with the Berry curvature. The group velocity is
\bqa && {\bm v}_F = {\bm \nabla}_{\bm p} \Big[ \Big( 1 + \frac{e}{c} \bm{\mathcal{B}}_{\bm p} \cdot {\bm B} \Big) |{\bm p}|  \Big] \Big|_{\bm p = \bm p_F} . \eqa
Solutions are given by
\bqa
\dot{\bm r}
&=& G \Big[ {\bm v}_F + e{\bm E} \times \bm{\mathcal{B}}_F + \frac{e}{c} \big( {\bm v}_F \cdot \bm{\mathcal{B}}_F \big) {\bm B} \Big] , \nn
\dot{\bm p}
&=& G \Big[ e{\bm E} + \frac{e}{c} {\bm v}_F \times {\bm B} + \frac{e^2}{c} \big( {\bm E} \cdot {\bm B} \big) \bm{\mathcal{B}}_F \Big], \eqa
with the measure-correction factor for the phase-space volume
\bqa G = \Big( 1 + \frac{e}{c} \bm{\mathcal{B}}_F \cdot {\bm B} \Big)^{-1} . \eqa
Recently, we find an interesting correction to this topologically modified Drude model, originating from an emergent Berry electric field that appears when the Berry magnetic field changes as a function of time \cite{Berry_Electric_Field}.

\subsection{Current conservation law in the topological Fermi-liquid theory}

Based on the topologically modified Boltzmann transport theory, it is straightforward to find the current conservation law
\bqa &&
\partial_{t} \rho_{\chi}(\bm{r},t) + \bm{\nabla}_{\bm{r}} \cdot \bm{j}_{\chi}(\bm{r},t) = \frac{k_{\chi}}{4\pi^{2}} \frac{e^2}{c} \bm{E} \cdot \bm{B}.
\eqa
Here, the ``conserved" density is
\bqa && \rho_{\chi}(\bm{r},t) = \int\! \frac{d^{3} \bm{p}}{(2\pi)^{3}} \ G_{\chi}^{-1} n_{\chi}(\bm{p};\bm{r},t) , \eqa
and the ``conserved" current is
\begin{widetext}
\bqa && \bm{j}_{\chi}(\bm{r},t) = \int\! \frac{d^{3} \bm{p}}{(2\pi)^{3}} \ \bigg\{ \Big( \bm{v}_{F}^{\chi} + e {\bm E} \times \boldsymbol{\mathcal{B}}_F^\chi + \frac{e}{c} (\bm{\mathcal{B}}_{F}^{\chi} \cdot \bm{v}_{F}^{\chi}) \bm{B} \Big) n_{\chi}(\bm{p};\bm{r},t) \nn &&\hspace{100pt} + [ - \partial_{\varepsilon} n_{eq}(\varepsilon) ] \sum_{\bm{p}'} \sum_{\chi' = \pm} F_{\chi\chi'}(\bm{p},\bm{p}') \Big( \bm{v}_{F}^{\chi} + \frac{e}{c} (\bm{\mathcal{B}}_{F}^{\chi} \cdot \bm{v}_{F}^{\chi}) \bm{B} \Big) \delta n_{\chi'}(\bm{p}';\bm{r},t) \bigg\} . \eqa
\end{widetext}
There exists an interaction-induced term. This term can be understood to originate from a back-flow current due to forward scattering, filling the empty space where electrons move away \cite{Boltzmann_LFL}. In Weyl metals, the group velocity is modified to $\bm{v}_{F}^{\chi} + \frac{e}{c} (\bm{\mathcal{B}}_{F}^{\chi} \cdot \bm{v}_{F}^{\chi}) \bm{B}$, involved with the effective dynamics of the emergent magnetic dipole moment under external magnetic fields.
\bqa && \frac{k_{\chi}}{4\pi^{2}} = \int\! \frac{d^{3} \bm{p}}{(2\pi)^{3}} \  \boldsymbol{\mathcal{B}}_F^{\chi} \cdot \bm{\nabla}_{\bm{p}} n_{\chi}(\bm{p};\bm{r},t) \eqa
is a magnetic charge in momentum space.

Resorting to the topologically modified Boltzmann transport theory, one can find dynamical response functions. We recall to sketch how to obtain the dynamical susceptibility in the Landau's Fermi-liquid state, for example. One may calculate the optical conductivity
\bqa &&
\sigma_{ij}(\bm{q},\nu) = \frac{1}{ i \nu } \frac{\sum_{\chi = \pm} \delta \bm{j}_{\chi}^{i}(\bm{q},\nu)}{\delta \bm{A}^{j}(\bm{q},\nu)}
\eqa
in interacting Weyl metals, turning on electric fields
\bqa && \bm{E} = - \frac{1}{c} \partial_{t} \bm{A}(\bm{r},t) . \eqa
In this study we focus on the derivation of a topological Fermi-liquid theory instead of its property.

\section{Derivation of a topological Fermi-liquid theory}

In order to derive the topological Fermi-liquid theory, we obtain the topological Fermi-gas theory first in the absence of interactions. A novel ingredient is that this derivation is based on the second quantization. We recall that the topologically modified Drude model has been derived within the first quantization. Then, we introduce all possible interactions between low energy fermions on the pair of chiral Fermi surfaces. Performing Shankar's renormalization group analysis, we find that the topological Fermi-liquid state is realized above the critical temperature of BCS instability, where only forward scattering is marginal as the Landau's Fermi-liquid state.

\subsection{Derivation of a topological Fermi-gas theory in the second quantization approach}

\subsubsection{Integration of high-energy fermions}

We start from a massless Dirac Lagrangian with a background chiral gauge field, given by
\bqa
\mathcal{L}_{WM} = \bar\psi(x) \Big( i\gamma^\alpha D_\alpha + \mu \gamma^0 + c_\alpha \gamma^\alpha \gamma^5 \Big) \psi(x) , \label{Dirac_Lagrangian_Chiral_Gauge_Field}
\eqa
which is fermionic part of Eq. (\ref{eq:TFL_action}) without interaction between fermions, and $m=0$.
As before, $\psi(x)$ is a four-component Dirac spinor with
\bqa && \bar\psi(x) = \psi^\dagger(x) \gamma^0 . \eqa
$\gamma^{\mu}$ is a four-by-four Dirac gamma matrix to satisfy the Clifford algebra
\bqa && \{ \gamma^\alpha, \gamma^\beta \} = 2\eta^{\alpha\beta} , \eqa
where the flat metric is
\bqa && \eta^{\alpha\beta} = \mathrm{diag}(1,-1,-1,-1) . \eqa
More explicitly, we have
\bqa && \gamma^0 = \begin{pmatrix} 0 & I \\ I & 0 \end{pmatrix}, ~~~~~ \gamma^i = \begin{pmatrix} 0 & -\sigma^i \\ \sigma^i & 0 \end{pmatrix} . \eqa
$D_\alpha = \partial_\alpha + i A_\alpha$ is the covariant derivative with an electromagnetic vector potential. For convenience, we use a unit with $e=c=1$ in this section.
\bqa && c_\alpha \equiv (c_0, {\bm c}) = (0, g {\bm B}) \eqa is the background chiral gauge field, given by the external magnetic field $\bm{B}$.
\bqa && \gamma^5 = i\gamma^0\gamma^1\gamma^2\gamma^3 = \begin{pmatrix} I & 0 \\ 0 & -I \end{pmatrix} \eqa
is the chiral matrix. $\mu$ is the chemical potential.

%
%
\begin{figure}
\includegraphics[width=8.5cm]{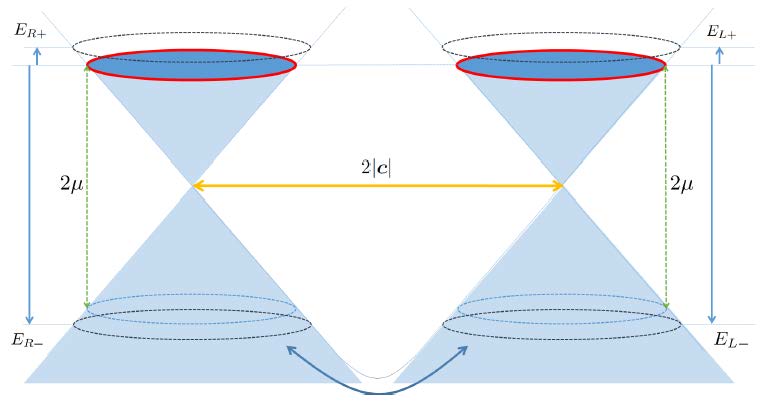}
\caption{Spectrum of Weyl electrons at $\mu \neq 0$.} \label{fig:Spectrum}
\end{figure}
%
%
It is straightforward to obtain the Dirac equation from the above Lagrangian as follows
\bqa
i \partial_{t} \psi(x) &=& \Big( -i\gamma^0 \gamma^i \partial_i -\mu - c_i \gamma^0 \gamma^i \gamma^5 \Big) \psi(x) \nn &\equiv& H_{WM} \psi(x) ,
\eqa
where $H_{WM}$ is an effective Hamiltonian for a Weyl metal phase. Here, we dropped $A_\alpha$ but retained $c_\alpha$. The Weyl band structure is given by
\bqa E_{\chi \pm} ({\bm p}) &=& \pm |{\bm p} + \chi {\bm c}| - \mu , \eqa
where $\chi = R \equiv 1$ and $\chi = L \equiv -1$ are the chirality of each Weyl band. See Fig. \ref{fig:Spectrum}. The corresponding eigenstates are
\bqa u_{R \pm} ({\bm p}) = \begin{pmatrix} \xi^{\pm}_{{\bm p} + {\bm c}} \\ 0 \end{pmatrix} , \ \ \  u_{L \pm} ({\bm p}) = \begin{pmatrix} 0 \\ \xi^{\mp}_{{\bm p} - {\bm c}} \end{pmatrix} , \eqa
respectively, for each chirality. Here, the two-component Weyl spinor is defined as follows
\bqa && {\bm v} \cdot {\bm \sigma} \xi_{\bm v}^{\pm} = \pm|{\bm v}| \xi_{\bm v}^{\pm} . \eqa

Our Fermi surfaces consist of two spheres of the radius $|\mu|$, and they are separated by the distance $2|{\bm c}|$ along the direction of ${\bm c}$ in momentum space (Fig. \ref{fig:Spectrum}). Since electron states of the Fermi surface centered at $-{\bm c}$ ($+{\bm c}$) have the right- (left-) chirality, the Fermi surface is referred to as a right- (left-) chiral Fermi surface. We note that the spin direction in the right- (left-) chiral Fermi surface is in parallel (anti-parallel) with ${\bm p}+{\bm c}$ (${\bm p} - {\bm c}$) for $\mu >0$, and opposite for $\mu <0$.

Assume that, for definiteness, the chemical potential $\mu$ is finite and positive. In the low-energy limit, much smaller than the Fermi energy, we expect that only low energy electrons near the Fermi surface are involved with physical responses. However, the equation of motion shows that these states are coupled to high energy modes with the energy near $-2\mu$, not allowing us to take into account such low energy fermions only. It is desirable to find a low-energy effective theory, integrating out such high-energy modes.

%
%
\begin{figure}
\includegraphics[width=8.5cm]{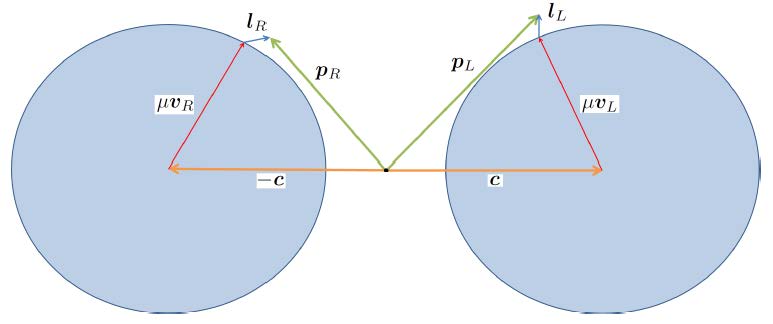}
\caption{A pair of chiral Fermi surfaces} \label{fig:ChiralFS}
\end{figure}
%
%

%
%
\begin{figure}
\centering \includegraphics[width=6cm]{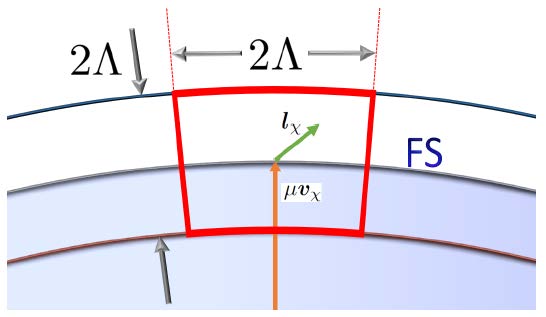}
\caption{Patch construction for an effective field theory. Here, $\Lambda$ represents the size of a patch (cutoff), characterized by a vector $\mu \bm{v}_{\chi}$ with a chemical potential $\mu$, where $\chi$ means the chirality of chiral Fermi surface (FS). $\bm{l}_{\chi}$ is a momentum to describe dynamics of a spinless chiral fermion.} \label{fig:Patch_Construction}
\end{figure}
%
%

In order to perform this procedure, it is convenient to consider a patch construction with a thin shell of thickness $\Lambda$ as follows (Figs. \ref{fig:ChiralFS} and \ref{fig:Patch_Construction}): Divide the thin shell near the Fermi surface into small patches of the dimension $\Lambda$ (the area and the thickness are order of $\Lambda$). Label the patch as a vector $\mu{\bm v}_\chi$ with $|{\bm v}_\chi|=1$, the position of the patch from the center of the chiral Fermi surface. The residual momentum within the patch is denoted by ${\bm l}_\chi$. In other words, we express the four momentum as
\bqa
p^\alpha_\chi = - \chi c^\alpha + \mu v^\alpha_\chi + l^\alpha_\chi .
\eqa
Here, $- \chi c^\alpha$ represents a Weyl point of the chirality $\chi$, enclosed by a chiral Fermi surface. See Fig. \ref{fig:ChiralFS}. $\mu v^\alpha_\chi$ with the chemical potential $\mu$ and $v^\alpha_\chi = (0, {\bm v}_\chi)$ denotes a patch position of the chirality $\chi$ on the chiral Fermi surface, where $|{\bm v}_\chi| = 1$.
\bqa |l^\alpha_\chi| < \Lambda \ll \mu \eqa
describes the dynamics of low-energy chiral fermions in the patch. See Fig. \ref{fig:Patch_Construction}. As a result, we decompose the Dirac spinor field as follows
\bqa
&&\psi(x) = \sum_\chi \sum_{{\bm v}_\chi} \int_{|l_\chi|<\Lambda}\!\frac{d^4l_\chi}{(2\pi)^4} \ e^{-i\chi {\bm c} \cdot {\bm x} + i\mu {\bm v}_\chi \cdot {\bm x} - il_\chi \cdot x} \nn
&& \Big[ Q_{\chi +}({\bm v}_\chi, l_\chi) u_{\chi +}({\bm p}) + Q_{\chi -}({\bm v}_\chi, l_\chi) u_{\chi -}({\bm p}) \Big] .
\eqa
Here, $Q_{\chi +}({\bm v}_\chi, l_\chi)$ is an amplitude (annihilation operator) of an eigenstate $u_{\chi +}({\bm p})$, which describes a positive energy band. On the other hand, $Q_{\chi -}({\bm v}_\chi, l_\chi)$ is an amplitude (annihilation operator) of an eigenstate $u_{\chi -}({\bm p})$, which describes a negative energy band. Taking into account the chiral fermion field
\bqa
&& q_{\chi\pm} ({\bm v}_\chi,x) = \int_{|l_\chi|<\Lambda}\!\frac{d^4l_\chi}{(2\pi)^4} \ e^{- il^\alpha_\chi x_\alpha} Q_{\chi \pm}({\bm v}_\chi, l_\chi) u_{\chi \pm}({\bm p}) , \nn
\eqa
defined within the patch of ${\bm v}_\chi$, we rewrite the Dirac spinor field as follows
\bqa
&& \psi_\chi(x) = \sum_{{\bm v}_\chi} e^{-i\chi {\bm c} \cdot {\bm x} + i\mu {\bm v}_\chi \cdot {\bm x}} \big[ q_{\chi +}({\bm v}_\chi, x) + q_{\chi -}({\bm v}_\chi, x) \big] , \nn \label{Decomposition_Dirac_Spinor}
\eqa
where
\bqa && \psi(x) = \sum_{\chi} \psi_\chi(x) . \eqa
This completes the decomposition of the Dirac spinor field in the patch construction.

Introducing Eq. (\ref{Decomposition_Dirac_Spinor}) into Eq. (\ref{Dirac_Lagrangian_Chiral_Gauge_Field}), we rewrite the Dirac Lagrangian with the chiral gauge field as follows
\bqa && \mathcal{L}_{WM} = \sum_\chi \sum_{{\bm v}_\chi} \Big[ \bar q_{\chi +} \gamma^0 \big( V_\chi \cdot i D \big) q_{\chi +} \nn && + \bar q_{\chi -} \big( 2\mu + \bar V_\chi \cdot iD \big) q_{\chi -} + \bar q_{\chi +} \big( \gamma_{\perp\chi} \cdot i D \big) q_{\chi -} \nn && + \bar q_{\chi -} \big( \gamma_{\perp \chi} \cdot i D \big) q_{\chi +} \Big] .
\eqa
Here, $\sum_{{\bm v}_\chi}$ and $\sum_\chi$ represent the summation for different patches and chiralities, respectively. The four-component spinor $q_{\chi \pm} = q_{\chi\pm} ({\bm v}_\chi,x)$ describe high ($+$) and low ($-$) energy fermion fields, respectively, of the chirality $\chi$ and the patch index ${\bm v}_\chi$. We also introduced
\bqa && \bar q_{\chi\pm} = q^\dagger_{\chi\pm} \gamma^0 \eqa
into the above expression. The velocity of each chiral fermion is defined as
\bqa && V_\chi^\alpha = (1, {\bm v}_\chi) , ~~~~~ \bar V_\chi^\alpha = (1, -{\bm v}_{\chi}) . \eqa
We note the following notation for the gamma matrix
\bqa && \gamma_{\perp \chi}^\alpha = \gamma^\alpha - \gamma_{\parallel\chi}^\alpha , \eqa
where a projected gamma matrix is given by
\bqa && \gamma_{\parallel\chi}^\alpha = (\gamma^0, {\bm v}_\chi ({\bm v}_\chi \cdot {\bm \gamma})) . \eqa
We recall that the covariant derivative works in the Fourier-transformed coordinate of the relative momentum $l_\chi$, defined at the chirality $\chi$ and the patch index ${\bm v}_\chi$.

We are ready to perform the gaussian integral for high-energy chiral fermions as follows
\bqa
Z &=& \int D q_{\chi +} D q_{\chi -} \ e^{i\int\!d^4x \ \mathcal{L}_{WM}}
\nn &=& \int D q_{\chi +} \ e^{i\int\!d^4x \ \mathcal{L}_{WM}^{eff}} .
\eqa
The effective Weyl metal Lagrangian is given by
\bqa
\mathcal{L}_{WM}^{eff} &=& \sum_\chi \sum_{{\bm v}_\chi} \Big[ q_{\chi +}^\dagger \big( V_\chi \cdot iD \big) q_{\chi +} \nn &+& \frac{1}{2\mu} q_{\chi +}^\dagger \big( \gamma_{\perp \chi} \cdot iD \big)^2 q_{\chi +} \nn
&+& \frac{1}{4\mu^2} q_{\chi +}^\dagger \big( \gamma_{\perp \chi} \cdot iD \big) \big( \bar V_\chi \cdot iD \big) \big( \gamma_{\perp \chi} \cdot iD \big) q_{\chi +} \Big] \nn &+& \mathcal{O}\left( \frac{1}{\mu^3} \right)
\eqa
up to the second order in the chemical potential. We recall that the chemical potential is the largest energy scale, regarded to be an expansion parameter.

\subsubsection{$1/\mu$ expansion}

The first order in the $1/\mu$ expansion is
\bqa &&
\mathcal{L}_{\mathcal{O}(\mu^{-1})} = \sum_{\chi} \sum_{\bm{v}_\chi} \frac{1}{2\mu} q_{\chi+}^\dagger ( \gamma_{\chi \perp}^\mu i D_\mu )^2 q_{\chi+} .
\eqa
Here, we have
\bqa
( \gamma_{\chi \perp}^{\mu} i D_{\mu} )^2 = \mathrm{diag}\Big[ (i D_{\chi \perp})^{2} - ( \bm{v}_\chi \cdot \bm{B} ) ( \bm{v}_\chi \cdot \bm{\sigma}) \Big] .
\eqa
Recalling
\bqa && q_{\chi+}^\dagger \bm{\sigma} q_{\chi+} = \chi \bm{v}_\chi q_{\chi+}^\dagger q_{\chi+} , \eqa
we obtain
\bqa &&
\mathcal{L}_{\mathcal{O}(\mu^{-1})} = \sum_\chi \sum_{ \bm{v}_\chi } \frac{1}{2 \mu} q_{\chi+}^\dagger [ ( i D_{\chi \perp} )^2 - \chi \bm{v} \cdot \bm{B} ] q_{\chi+} . \nn
\eqa
where a chirality-dependent Zeeman-type shift on each Fermi surface appears to be identified with the term of $|\bm{p}| \bm{\mathcal{B}}_F \cdot \bm{B}$, which modifies one-particle energy due to the Berry curvature in the presence of electromagnetic fields. This modification has been shown to occur from a nontrivial manifestation of the Lorentz symmetry, interpreted as a coupling term between an emergent magnetic dipole moment and an external magnetic field \cite{CME4,CME6,Berry_Fermi_Liquid_Theory,Magnetic_Moment_Lorentz_Symmetry1,Magnetic_Moment_Lorentz_Symmetry2}. There exist irrelevant corrections in the dispersion relation, associated with the tangential momentum
\bqa \bm{l}_{\perp \chi} = \bm{l}_{\chi} - \bm{v}_\chi ( \bm{v}_\chi \cdot \bm{l}_\chi ) , \eqa
as expected from the dimensional analysis. This term describes the curvature effect of the Fermi surface.

All terms in $\mathcal{O}(\mu^{-2})$ are found to be
\bqa
&& \mathcal{L}_{\mathcal{O}(\mu^{-2})} \nn && = - \sum_{\chi} \sum_{\bm{v}_\chi} \frac{1}{4 \mu^2} q_{\chi+}^\dagger ( \gamma_{\chi \perp}^\mu i D_\mu ) (\bar{V}_\chi^\nu i D_\nu ) ( \gamma_{\chi \perp}^\rho i D_\rho ) q_{\chi+} . \nn
\eqa
We rearrange this term as follows
\bqa
&& (\gamma_{\chi \perp}^\mu i D_\mu ) ( \bar{V}_\chi^\nu i D_\nu ) ( \gamma_{\chi \perp}^\rho i D_\rho ) \nn && = \gamma_{\chi \perp}^i \gamma_{\chi \perp}^j \bar{V}_\chi^\mu (-i F_{i \mu} ) ( i D_j ) + ( \bar{V}_\chi^\mu i D_\mu ) ( \gamma_{\chi \perp}^\nu i D_\nu )^2 . \nn
\eqa
Resorting to
\bqa
&& \gamma_{\chi \perp}^i \gamma_{\chi \perp}^j \nn && = \begin{pmatrix} - \delta^{ij} + i \epsilon^{ijk} v_\chi^k + v_\chi^i v_\chi^j & 0 \\ 0 & - \delta^{ij} - i \epsilon^{ijk} v_\chi^k + v_\chi^i v_\chi^j \end{pmatrix} , \nn
\eqa
we find
\begin{widetext}
\bqa
&& \gamma_{\chi \perp}^{i} \gamma_{\chi \perp}^{j} \bar{V}_\chi^\mu (-i F_{i \mu} ) ( i D_j ) \nn
&& =  i \bm{E} \cdot i \bm{D} - i ( \bm{v} \cdot \bm{E} ) ( \bm{v} \cdot i \bm{D} ) - \chi ( \bm{v} \times \bm{E} ) \cdot i \bm{D} - i ( \bm{v} \times \bm{B} ) \cdot i \bm{D} + \chi ( \bm{v} \cdot \bm{B} ) ( \bm{v} \cdot i \bm{D} ) - \chi \bm{B} \cdot i \bm{D}
\eqa
and
\bqa
&& ( \bar{V}^{\mu} i D_{\mu} ) ( \gamma_{\perp}^{\nu} i D_{\nu} )^2 \nn
&& = g_2 (\bm{l}_\chi, \omega_\chi, \bm{A} ) - 2 i \bm{E} \cdot i \bm{D} + 2i ( \bm{v} \cdot \bm{E} )(\bm{v} \cdot i \bm{D}) + i ( \bm{v}_\chi \times \bm{B} ) \cdot i \bm{D} - \chi (\bm{v}_\chi \cdot \bm{B} ) ( \bm{v}_\chi \cdot i \bm{D} ) ,
\eqa
where
\bqa
g_2 (\bm{l}_\chi, \omega_\chi, \bm{A} ) &=& \Big[ - | \bm{l}_\chi |^2 + 2 (\bm{l}_\chi \cdot \bm{A}) - | \bm{A} |^2 + ( \bm{v}_\chi \cdot \bm{l}_\chi - \bm{v}_\chi \cdot \bm{A} )^2 \Big] ( \bm{v}_\chi \cdot \bm{l}_\chi - \bm{v}_\chi \cdot \bm{A}) \nn && - | \bm{l}_\chi |^2 \omega_\chi + 2 \omega_\chi ( \bm{l}_{\chi} \cdot \bm{A} ) - | \bm{A} |^2 \omega_\chi .
\eqa
As a result, we reach the following expression
\bqa
\mathcal{L}_{\mathcal{O}(\mu^{-2})}
&=& - \sum_\chi \sum_{\bm{v}_\chi} \frac{1}{4 \mu^2} q_{\chi+}^\dagger \Big[ i ( \bm{v}_\chi \times \bm{E} ) \cdot ( \bm{v}_\chi \times i \bm{D} ) -  \chi (\bm{v}_\chi \times \bm{E} ) \cdot i \bm{D} - \chi \bm{B} \cdot i \bm{D} \Big] q_{\chi+} \nn
&+& (\text{higher-order dispersion terms}).
\eqa
\end{widetext}
There are three relevant terms. The first term, which has no chirality dependence, is well known to appear as a non-hermitian term in the massive limit of the Dirac theory \cite{Huang_QFT_Textbook} while other two terms, which have chirality dependence, correspond to those of the first quantization representation, involved with the Berry curvature \cite{CME3,CME4,CME5,CME6,CME7,Boltzmann_Chiral_Anomaly1,Boltzmann_Chiral_Anomaly2,Boltzmann_Chiral_Anomaly3,
Boltzmann_Chiral_Anomaly4,Boltzmann_Chiral_Anomaly5,Boltzmann_Chiral_Anomaly6}. The first non-Hermitian term reflects the fact that $q_{\chi+}^\dagger q_{\chi+}$ is not conserved due to the existence of $q_{\chi -}$ states via the hybridization. We neglect this term by requiring a consistent one-particle interpretation of the $q_{\chi +}$ state. The second term is involved with the anomalous Hall effect and the last term is associated with the chiral magnetic effect. We point out that the gauge invariance is satisfied naturally as it must be.

\subsubsection{An effective field theory for a topological Fermi-gas state}

Gathering all terms up to the second order in the $1/\mu$ expansion, where all terms have been rearranged to show effects of both the Berry curvature and chiral anomaly as effective potentials, we obtain an effective Lagrangian for a Weyl metal state
\bqa
&& \mathcal{L}_{WM}^{eff} = \mathcal{L}_{\mathcal{O}(\mu^{0})} + \mathcal{L}_{\mathcal{O}(\mu^{-1})} + \mathcal{L}_{\mathcal{O}(\mu^{-2})} .
\eqa
The zeroth order gives a typical form of the kinetic energy of low-energy fermions near the Fermi surface
\bqa
&& \mathcal{L}_{\mathcal{O}(\mu^{0})} = \sum_\chi \sum_{{\bm v}_\chi} q_{\chi +}^\dagger \big( V_\chi \cdot i D \big) q_{\chi +} . \label{TFL_Zeroth_Order}
\eqa
The first order results in a nontrivial correction to the dispersion relation due to the contribution from the magnetic dipole moment given by the Berry curvature, consistent with the correction of the first-quantization approach
\bqa
\mathcal{L}_{\mathcal{O}(\mu^{-1})} &=& \sum_\chi \sum_{{\bm v}_\chi} \frac{1}{2\mu} q_{\chi +}^\dagger \big( - \chi {\bm v}_\chi \cdot {\bm B} \big) q_{\chi +} \nn
&+& \text{(higher dispersion terms)} . \label{TFL_First_Order}
\eqa
The second order expansion gives rise to geometrically nontrivial contributions from the Berry curvature, involved with the anomalous Hall effect and the chiral magnetic effect
\bqa
\mathcal{L}_{\mathcal{O}(\mu^{-2})} &=& \sum_\chi \sum_{{\bm v}_\chi} \frac{\chi}{4\mu^2} q_{\chi +}^\dagger \Big[ \big( {\bm v}_\chi \times {\bm E} + {\bm B} \big) \cdot i{\bm D} \Big] q_{\chi +} \nn
&+& \text{(higher dispersion \& non-Hermitian terms)} . \nn \label{TFL_Second_Order}
\eqa

%
%

The next procedure is to enslave spin degrees of freedom in the dynamics of Weyl electrons, which may be regarded to originate from the spin-momentum locking, given by
\bqa
q_{R+}({\bm v}_R, x) = U_{\bm p}^R \begin{pmatrix} f_R({\bm v}_R, x) \\ 0 \\ 0 \\ 0 \end{pmatrix} \label{Spin_Enslavement_Positive_Chirality}
\eqa
and
\bqa
q_{L+}({\bm v}_L, x) = U_{\bm p}^L \begin{pmatrix} 0 \\ 0 \\  0 \\ f_L({\bm v}_L, x) \end{pmatrix} , \label{Spin_Enslavement_Negative_Chirality}
\eqa
respectively. Here, $U_{\bm p}^\chi$ is a unitary matrix that makes $\xi_{{\bm p} + \chi {\bm c}}^{\pm}$ be diagonalized. This procedure is called spin enslavement \cite{Magnetic_Moment_Lorentz_Symmetry1}. In terms of $f_\chi$, the effective action now reads
\bqa
\mathcal{L}_{WM}^{eff} &=& \sum_\chi \sum_{{\bm v}_\chi} f_\chi^\dagger \bigg[ V_\chi \cdot i D - \frac{1}{2\mu} {\bm v}_\chi \cdot {\bm B} \nn &+& \frac{\chi}{4\mu^2} \big( {\bm v}_\chi \times {\bm E} + {\bm B} \big) \cdot i{\bm D} \bigg] f_\chi \nn &+& \text{(higher dispersion \& non-hermitian terms)} . \nn \label{WM_Effective_Action_Derivation}
\eqa

Based on this derivation and reflecting the semiclassical action in the first quantization expression, we propose a topological Fermi-gas theory for the dynamics of spinless chiral Fermions on a pair of chiral Fermi surfaces (in the imaginary-time formalism)
\bqa
&& Z_{TFG} = \int D f_\chi \ \exp [-\mathcal{S}_{TFG}] ,
\eqa
where an effective action is derived as follows
\bqa
\mathcal{S}_{TFG} &=& \int_0^\beta \!d\tau \int d^3\bm x \ f_{\chi}^\dagger ({\bm x},\tau) \bigg[ \partial_\tau - \varphi \nn
&+& G_3^{\chi -1} \Big\{ \dot {\bm x}_F^\chi \cdot i\big( {\bm \nabla}_{\bm x} + i{\bm A} \big) - \dot {\bm p}_F^\chi \cdot \bm{\mathcal{A}}_F^\chi \Big\} \bigg] f_{\chi}({\bm x},\tau) . \label{eq:TFGT} \nn
\eqa
Here,
\bqa && G_3^{\chi} = (1+ \bm{\mathcal{B}}_F^\chi \cdot {\bm B})^{-1} \eqa
is the phase-space volume normalization factor and the subscript of F is the Fermi momentum, which appears from the chemical potential in Eq. (\ref{WM_Effective_Action_Derivation}).

Although this effective field theory looks normal at a glance, the dispersion of such spinless fermions is seriously modified, where the velocity of the dispersion is renormalized as follows
\bqa
\dot {\bm x}_F^\chi
&=& G^\chi_3 \Big[ {\bm v}_F^\chi + {\bm E} \times \boldsymbol{\mathcal{B}}_F^\chi + \big( \boldsymbol{\mathcal{B}}_F^\chi \cdot {\bm v}_F^\chi \big) {\bm B} \Big] , \label{Drude_Model_Group_Velocity}
\eqa
where $\bm{v}_{F}^{\chi}$ is the Fermi velocity and $\bm{\mathcal{B}}_{F}^{\chi}$ is the Berry curvature on the chiral Fermi surface, satisfying $\bm{\nabla}_{\bm{p}} \cdot \bm{\mathcal{B}}_{\bm{p}}^{\chi} = 2\pi\chi \delta^{(3)}(\bm{p} - \chi \bm{c})$. We emphasize that chemical-potential and chirality dependent terms are just translated into Berry-curvature terms for more general expressions from Eq. (\ref{WM_Effective_Action_Derivation}) to Eq. (\ref{eq:TFGT}). We recall $\boldsymbol{\mathcal{B}}_F^\chi = \chi \frac{{\bm v}_\chi}{2\mu^2}$ in Eq. (\ref{WM_Effective_Action_Derivation}). This dispersion relation with non-minimal gauge couplings may be regarded to be an essential feature for the low-energy dynamics of chiral fermions on a pair of chiral Fermi surfaces, derived from from QED$_{4}$ with a spatially inhomogeneous axion term. This expression turns out to be identical to the group velocity, derived from the wave-packet picture \cite{Berry_Curvature_Review1,Berry_Curvature_Review2} and the first-quantized Weyl Lagrangian \cite{CME4,CME6}, where the second term is responsible for anomalous Hall effect and the third term is the source of chiral magnetic effect, both of which are derived in the $1/\mu^{2}$ order.

In addition to the velocity renormalization, there appears an effective potential associated with the Berry-phase term, where the Drude model with the Lorentz force is modified as follows
\bqa
\dot {\bm p}_F^\chi
&=& G^\chi_3 \Big[ {\bm E} + {\bm v}_F^\chi \times {\bm B} + \big( {\bm E} \cdot {\bm B} \big) \boldsymbol{\mathcal{B}}_F^\chi \Big] . \label{Drude_Model_Lorentz_Force}
\eqa
We would like to point out that the Berry gauge-field dependent term of $\boldsymbol{\dot{p}}_{F}^{\chi} \cdot \bm{\mathcal{A}}_{F}^{\chi}$ has not been derived from QED$_{4}$ with the spatially inhomogeneous axion term, but introduced into Eq. (\ref{eq:TFGT}) explicitly as our ansatz. Recalling the first-quantization approach, this term is expected to result from $i \boldsymbol{\dot{p}}_{F}^{\chi} \cdot \nabla_{\bm{p}}\big|_{\bm{p}_{F}}$ when the spin-enslavement procedure is taken into account. Although we cannot figure out why we fail to derive this term at present, we consider such a term based on physical reasonability. The $\bm{E} \cdot \bm{\mathcal{A}}_{F}^{\chi}$ term is analogous to $\bm{E} \cdot \bm{P}$, where $\bm{P}$ is an electric polarization density associated with permanent dipole moments, here originating from the Berry gauge field. This reminds us of the Resta's geometric mechanism for electric polarization \cite{Resta_Polarization}. The $(\bm{v}_{F}^{\chi} \times \bm{B}) \cdot \bm{\mathcal{A}}_{F}^{\chi} = - (\boldsymbol{v}_{F}^{\chi} \times \bm{\mathcal{A}}_{F}^{\chi}) \cdot \boldsymbol{B}$ term is analogous to $\bm{L} \cdot \bm{B}$ with an angular momentum $\bm{L}$, where an anomalous displacement given by the Berry gauge field is responsible for the anomalous angular momentum. These two terms can arise in the $1/\mu$ order. The third term reflects the chiral anomaly on the chiral Fermi surface in the $1/\mu^{3}$ order. Now, the topological Fermi-gas theory Eq. (\ref{eq:TFGT}) shows parallelism with the first-quantization approach.

\subsection{Derivation of Boltzmann transport theory for a topological Fermi-gas state}

Although the topological Fermi-gas theory looks naturally connected with the path-integral formulation of the first quantization, we confirm Eq. (\ref{eq:TFGT}) [or Eq. (\ref{WM_Effective_Action_Derivation})] with Eqs. (\ref{Drude_Model_Group_Velocity}) and (\ref{Drude_Model_Lorentz_Force}), deriving a topologically modified Boltzmann equation in the collisionless limit. We will follow the standard procedure: Derive the equation-of-motion of a lesser Green's function in terms of the center-of-mass coordinate and the relative coordinate \cite{Mahan_Textbook}.

The equation-of-motion for $f_\chi$ is given by
\bqa && \delta \mathcal{S}_{WM}^{eff} \Big/ \delta f_\chi^\dagger = 0 , \eqa
resulting in
\begin{widetext}
\bqa
&& \Bigg[ i\partial_t - A_t - \Big\{ {\bm v}_\chi - \frac{\chi}{4\mu^2} \big( {\bm v}_\chi \times {\bm E} + {\bm B} \big) \Big\} \cdot \big( i{\bm \nabla} - {\bm A} \big) - \frac{\chi}{2\mu} {\bm v}_\chi \cdot {\bm B} - \frac{1}{4\mu^2} i\big( {\bm v}_\chi \times {\bm E} \big) \cdot \big\{ {\bm v}_\chi \times (i{\bm \nabla} - {\bm A}) \big\} \nn
&& + \frac{1}{2\mu} g_1^\chi({\bm x}, {\bm A}) - \frac{1}{4\mu^2} g_2^\chi({\bm x}, t, {\bm A}) \Bigg] f_{\chi}(x) = 0 . \label{eq:EOM_Eff_WM}
\eqa
We recall that the group velocity is modified as ${\bm v}_\chi - \frac{\chi}{4\mu^2} \big( {\bm v}_\chi \times {\bm E} + {\bm B} \big)$, discussed before. The coupling term between the magnetic dipole moment and the applied magnetic field is seen by $- \frac{\chi}{2\mu} {\bm v}_\chi \cdot {\bm B}$. Here, we include all terms up to $\mathcal{O}(1/\mu^2)$, where $- \frac{1}{4\mu^2} i\big( {\bm v}_\chi \times {\bm E} \big) \cdot \big\{ {\bm v}_\chi \times (i{\bm \nabla} - {\bm A}) \big\}$ originates from the non-hermitian term and
\bqa
g_1^{\chi} (\bm{l}_\chi , \bm{A}) & = & - |\bm{l}_\chi -\bm{A}|^2 + (\bm{v}_\chi \cdot \bm{l}_\chi - \bm{v}_\chi \cdot \bm{A} )^2 , \nn
g_2^{\chi} (\bm{l}_\chi, \omega_\chi, \bm{A} ) & = & \Big[- | \bm{l}_\chi |^2 + 2 (\bm{l}_\chi \cdot \bm{A}) - | \bm{A} |^2 + ( \bm{v}_\chi \cdot \bm{l}_\chi - \bm{v}_\chi \cdot \bm{A} )^2 \ \Big] ( \bm{v}_\chi \cdot \bm{l}_\chi - \bm{v}_\chi \cdot \bm{A}) \nn &-& | \bm{l}_\chi |^2 \omega_\chi + 2 \omega_\chi ( \bm{l}_{\chi} \cdot \bm{A} ) - | \bm{A} |^2 \omega_\chi ,
\eqa
\end{widetext}
are Fourier-transformed functions to describe irrelevant corrections in the dispersion relation.

The lesser Green's function is defined by
\bqa
G_\chi^< (x_1, x_2) = i \left< f_\chi^\dagger(x_2) f_\chi(x_1) \right> .
\eqa
Differentiating $G_\chi^<(x_1, x_2)$ with respect to $\partial_{t_1}$ and $\partial_{t_2}$, and using Eq. (\ref{eq:EOM_Eff_WM}) and its hermitian conjugate, we obtain the following equation-of-motion of $G_\chi^<(x_1, x_2)$
\begin{widetext}
\bqa
&& \Bigg[ i(\partial_{t_1} + \partial_{t_2}) - A_t(x_1) + A_t(x_2) - \Big\{ {\bm v}_\chi - \frac{\chi}{4\mu^2} \big( {\bm v}_\chi \times {\bm E} + {\bm B} \big) \Big\} \cdot \big\{ i{\bm \nabla}_{x_1} + i{\bm \nabla}_{x_2} - {\bm A}(x_1) + {\bm A}(x_2) \big\} \nn &&- \frac{1}{4\mu^2} i \big( {\bm v}_\chi \times {\bm E} \big) \cdot \Big({\bm v}_\chi \times \big\{ i{\bm \nabla}_{x_1} - i{\bm \nabla}_{x_2} - {\bm A}(x_1) - {\bm A}(x_2) \big\} \Big) + \frac{1}{2\mu} \big\{ g_1^\chi({\bm x}_1, {\bm A}(x_1)) - g_1^\chi({\bm x}_2, {\bm A}(x_2)) \big\} \nn && - \frac{1}{4\mu^2} \big\{ g_2^\chi({\bm x}_1, t_1, {\bm A}(x_1)) - g_2^\chi({\bm x}_2, t_2, {\bm A}(x_2)) \big\} \Bigg] G_\chi^<(x_1, x_2) = 0 .
\eqa
\end{widetext}
Introducing the center-of-mass and relative coordinates
\bqa
&&({\bm R}, T) = \frac{1}{2} (x_1^\alpha + x_2^\alpha)
\eqa
and
\bqa
({\bm r}, t) = x_1^\alpha - x_2^\alpha ,
\eqa
respectively, and performing the Fourier transformation for the relative coordinate
\bqa && -i{\bm \nabla}_r \rightarrow {\bm l} , ~~~~~ {\bm r} \rightarrow -i{\bm \nabla}_l, ~~~~~ t \rightarrow i\partial_\omega , \eqa
we arrive at the quantum Boltzmann equation in the collisionless limit,
\begin{widetext}
\bqa
&&\Bigg[ i\frac{\partial}{\partial T} + \Big\{ {\bm v}_\chi - \frac{\chi}{4\mu^2} \big( {\bm v}_\chi \times {\bm E} + {\bm B} \big) \Big\} \cdot i{\bm \nabla}_R
+ \Big\{ {\bm E} + \frac{1}{2} {\bm v}_\chi \times {\bm B} + \frac{\chi}{8\mu^2} ({\bm E} \cdot {\bm B}) {\bm v}_\chi - \frac{\chi}{8\mu^2} ({\bm v}_\chi \cdot {\bm B}) {\bm E} \Big\} \cdot i{\bm \nabla}_l \nn && - \frac{1}{4\mu^2} i \big( {\bm v}_\chi \times {\bm E} \big) \cdot \Big\{ {\bm v}_\chi \times {\bm l} + {\bm v}_\chi \times ({\bm R} \times {\bm B}) \Big\}
+ \frac{1}{2\mu} \big\{ g_1^\chi({\bm R} + {\bm r}/2, {\bm A}(R + r/2)) - g_1^\chi({\bm R} - {\bm r}/2, {\bm A}(R-r/2)) \big\} \nn && - \frac{1}{4\mu^2} \big\{ g_2^\chi(R+r/2, {\bm A}(R+r/2)) - g_2^\chi(R-r/2, {\bm A}(R-r/2)) \big\}
\Bigg] G_\chi^<({\bm l}, \omega; {\bm R},T) = 0 ,
\eqa
\end{widetext}
where Fourier transformed functions are used for $g_1^\chi$ and $g_2^\chi$, respectively. This Boltzmann equation is identical to the Boltzmann transport theory Eq. (\ref{Boltzmann_Transport_Theory_Weyl_Metal}) essentially, except for keeping both the non-hermitian term and higher-order corrections of the dispersion relation in the above expression and considering the contribution from forward scattering in Eq. (\ref{Boltzmann_Transport_Theory_Weyl_Metal}). In order to see this correspondence more explicitly, we rewrite the effective group velocity in $\dot{\bm{r}}_{\chi} \cdot i{\bm \nabla}_R$ and the effective Lorentz force in $\dot{\bm{p}}_{\chi} \cdot i{\bm \nabla}_l $ as follows
\bqa && \dot{\bm{r}}_{\chi} \equiv {\bm v}_\chi - \frac{\chi}{4\mu^2} \big( {\bm v}_\chi \times {\bm E} + {\bm B} \big) \nn && \approx \Big(1 + \frac{\chi}{8\mu^2} ({\bm v}_\chi \cdot {\bm B}) \Big)^{-1} \Big\{ {\bm v}_\chi - \frac{\chi}{4\mu^2} \big( {\bm v}_\chi \times {\bm E} + {\bm B} \big) \Big\} \nn && + \frac{\chi}{8\mu^2} ({\bm v}_\chi \cdot {\bm B}) {\bm v}_\chi \eqa
and
\bqa && \dot{\bm{p}}_{\chi} \equiv {\bm E} + \frac{1}{2} {\bm v}_\chi \times {\bm B} + \frac{\chi}{8\mu^2} ({\bm E} \cdot {\bm B}) {\bm v}_\chi - \frac{\chi}{8\mu^2} ({\bm v}_\chi \cdot {\bm B}) {\bm E} \nn && \approx \Big(1 + \frac{\chi}{8\mu^2} ({\bm v}_\chi \cdot {\bm B}) \Big)^{-1} \Big\{ {\bm E} + \frac{1}{2} {\bm v}_\chi \times {\bm B} + \frac{\chi}{8\mu^2} ({\bm E} \cdot {\bm B}) {\bm v}_\chi \Big\} \nn && + \frac{\chi}{16 \mu^2} ({\bm v}_\chi \cdot {\bm B}) {\bm v}_\chi \times {\bm B} , \eqa
respectively, where the phase-space volume measure has been introduced. Here, we keep all terms up to the $\mathcal{O}(\mu^{-2})$ order. It turns out that there exist additional terms of $\frac{\chi}{8\mu^2} ({\bm v}_\chi \cdot {\bm B}) {\bm v}_\chi$ and $\frac{\chi}{16 \mu^2} ({\bm v}_\chi \cdot {\bm B}) {\bm v}_\chi \times {\bm B}$ in these equations, respectively. Unfortunately, we do not understand the origin of this discrepancy.

If we repeat exactly the same procedure for the topological Fermi-gas field theory of Eq. (\ref{eq:TFGT}) with Eqs. (\ref{Drude_Model_Group_Velocity}) and (\ref{Drude_Model_Lorentz_Force}), we reach the following topological Boltzmann transport theory in the collisionless limit
\bqa
&& \Bigl( \frac{\partial}{\partial T} + \bm{\dot{x}}_{F}^{\chi} \cdot \bm{\nabla}_{\bm{R}} + \boldsymbol{\dot{p}}_{F}^{\chi} \cdot \bm{\nabla}_{\bm{p}} \Bigr) G_{\chi}^{<}(\bm{p};\bm{R},T) = 0 , \nn \label{Boltzmann_TRansport_Theory_Derivation_From_TFGT}
\eqa
where $\bm{l}$ has been replaced with $\bm{p}$. Here, the $\boldsymbol{\dot{p}}_{F}^{\chi} \cdot \bm{\nabla}_{\bm{p}}$ term has nothing to do with $- \dot {\bm p}_F^\chi \cdot \bm{\mathcal{A}}_F^\chi$ in the effective action of Eq. (\ref{eq:TFGT}). This expression is identical with the Boltzmann transport theory Eq. (\ref{Boltzmann_Transport_Theory_Weyl_Metal}) exactly, except for the contribution from forward scattering in Eq. (\ref{Boltzmann_Transport_Theory_Weyl_Metal}). In other words, the following effective action
\bqa
\mathcal{S}_{eff} &=& \int_0^\beta \!d\tau \int d^3x \ f_{\chi}^\dagger ({\bm x},\tau) \bigg[ \partial_\tau - \varphi \nn
&+& G_3^{\chi -1} \dot {\bm x}_F^\chi \cdot i\big( {\bm \nabla}_{\bm x} + i{\bm A} \big) \bigg] f_{\chi}({\bm x},\tau)
\eqa
gives rise to the Boltzmann transport theory of Eq. (\ref{Boltzmann_TRansport_Theory_Derivation_From_TFGT}). We would like to point out that this topological Boltzmann transport theory has been also derived in the collisionless limit, performing the similar coarse-graining procedure but from QED$_{4}$ directly \cite{Boltzmann_Chiral_Anomaly1}.

\subsection{Longitudinal positive magnetoconductivity}

Since the effective field theory of Eq. (\ref{eq:TFGT}) with Eqs. (\ref{Drude_Model_Group_Velocity}) and (\ref{Drude_Model_Lorentz_Force}) reproduces the topological Boltzmann equation and the topological Boltzmann transport theory predicts the enhancement of the longitudinal conductivity, it is natural to expect that the $B^{2}$ enhancement should be also predicted by the effective field theory within the Kubo formula. Taking derivatives twice with respect to the electromagnetic field in Eq. (\ref{eq:TFGT}), we obtain the current-current correlation function
\bqa
&& \Pi_{ij}(i\Omega) \nn && = \frac{1}{V}\sum_{\bm{p}} \Bigl( \boldsymbol{v}_{F}^{\chi} + \frac{e}{c} \boldsymbol{\mathcal{B}}_{F}^{\chi} \cdot \boldsymbol{v}_{F}^{\chi} \boldsymbol{B} \Bigr)_{i} \Bigl( \boldsymbol{v}_{F}^{\chi} + \frac{e}{c} \boldsymbol{\mathcal{B}}_{F}^{\chi} \cdot \boldsymbol{v}_{F}^{\chi} \boldsymbol{B} \Bigr)_{j} \nn
&& \frac{1}{\beta} \sum_{i\omega} G_{\chi}(\bm{p},i\omega+i\Omega) G_{\chi}(\bm{p},i\omega)
\end{eqnarray}
with the Green's function of
\bqa
G_{\chi}(\bm{p},i\omega) = \frac{1}{ i \omega + \bm{\dot{x}}_{F}^{\chi} \cdot \bm{p} + \boldsymbol{\dot{p}}_{F}^{\chi} \cdot \bm{\mathcal{A}}_{F}^{\chi} + i \gamma_{imp} \mbox{sgn}(\omega)} , \nn
\eqa
where $\gamma_{imp}$ is a scattering rate due to nonmagnetic randomness, introduced in the Born approximation \cite{Mahan_Textbook}. Here, the coupling constant $e$ and the speed of light $c$ is written explicitly. An essential point in this expression is that the group velocity is modified due to the chiral magnetic effect, given by $\boldsymbol{v}_{F}^{\chi} \longrightarrow \boldsymbol{v}_{F}^{\chi} + \frac{e}{c} \boldsymbol{\mathcal{B}}_{F}^{\chi} \cdot \boldsymbol{v}_{F}^{\chi} \boldsymbol{B}$.

The longitudinal magneto-conductivity along the applied magnetic field direction ($\bm{B} = B \bm{\hat{z}}$) is given by
\bqa
\sigma_{zz}
&& = -\lim_{\Omega\rightarrow 0} \mathrm{Im} \left[ \frac{1}{\Omega} \Pi_{zz}(i\Omega \rightarrow \Omega + i\delta) \right] \nn
&& = \frac{N_{F}}{4\pi^{2}} \int_{-1}^{1} d \cos \theta \int_{0}^{2\pi} d \varphi \int_{-\infty}^{\infty} d \xi \nn
&& \frac{\gamma_{imp}^{2} \Bigl( |\boldsymbol{v}_{F}^{\chi}|^{2} \cos^{2} \theta + \frac{e^{2}}{c^{2}} |\boldsymbol{\mathcal{B}}_{F}^{\chi}|^{2} |\boldsymbol{v}_{F}^{\chi}|^{2} B^{2} \Bigr)}{\Bigl\{\Bigl(\boldsymbol{\dot{p}}_{F}^{\chi} \cdot \bm{\mathcal{A}}_{F}^{\chi} + \bigl[1 + \frac{e}{c} |\boldsymbol{\mathcal{B}}_{F}^{\chi}| B \cos \theta \bigr] \xi \Bigr)^{2} + \gamma_{imp}^{2}\Bigr\}^{2}} \nn
&& \approx \Bigl( 1 + 12 \frac{e^{2}}{c^{2}} [|\bm{\mathcal{B}}_{F}^{R}|^{2} + |\bm{\mathcal{B}}_{F}^{L}|^{2}] B^{2} \Bigr) \sigma_{D} ,
\eqa
where
\bqa && \sigma_{D} = \frac{1}{24} N_{F} |\boldsymbol{v}_{F}|^{2} \gamma_{imp}^{-1} \eqa
is the Drude conductivity. We recover the $B^{2}$ enhancement for the longitudinal conductivity, which occurs when the electrical current is driven along the same direction of the applied magnetic field. A similar calculation shows that the longitudinal conductivity perpendicular to the applied magnetic field direction is $\sigma_{xx} = \sigma_{yy} = \sigma_D$, where the chiral magnetic effect does not exist, i.e., $\boldsymbol{v}_{F}^{\chi} + \frac{e}{c} \boldsymbol{\mathcal{B}}_{F}^{\chi} \cdot \boldsymbol{v}_{F}^{\chi} \boldsymbol{B} \longrightarrow \boldsymbol{v}_{F}^{\chi}$ in the above Kubo formula.

\subsection{Renormalization group analysis for effects of four-fermion interactions on a pair of chiral Fermi surfaces}
\label{subsection:RG}

Following the Landau's Fermi-liquid theory, we investigate the role of four-fermion interactions in the low-energy dynamics of chiral fermions on the pair of chiral Fermi surfaces. We recall such four-fermion interactions
\bqa
\mathcal{L}_{int}
&=& \lambda_s (\bar{\psi} \psi)^2 + \lambda_v (\bar{\psi} \gamma^\mu \psi)^2 + \lambda_{as} ( \bar{\psi} \gamma^{\mu \nu} \psi)^2 \nn
&+& \lambda_{pv} ( \bar{\psi} \gamma^\mu \gamma^5 \psi)^2 + \lambda_{ps} ( \bar{\psi} \gamma^5 \psi)^2 ,
\eqa
where $\psi$ is a four-component Dirac spinor. These four-fermion interactions are irrelevant in the case of zero chemical potential as long as their strengths remain weak. On the other hand, the presence of a Fermi surface gives rise to marginal interactions for some specific kinematic channels.

\begin{figure}
\includegraphics[width=8.5cm]{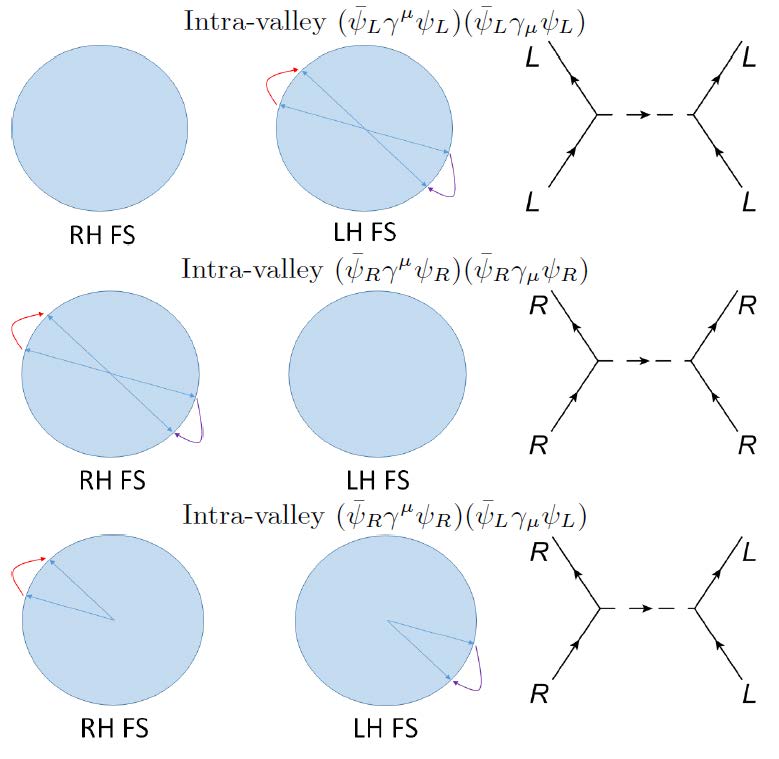}
\caption{All possible (Lorentz-invariant) four-fermion intra-valley scattering events on the pair of chiral Fermi surfaces from Eqs. (\ref{Interaction_decomposition_Intra_Valley}). } \label{intra}
\end{figure}

\begin{figure}
\includegraphics[width=8.5cm]{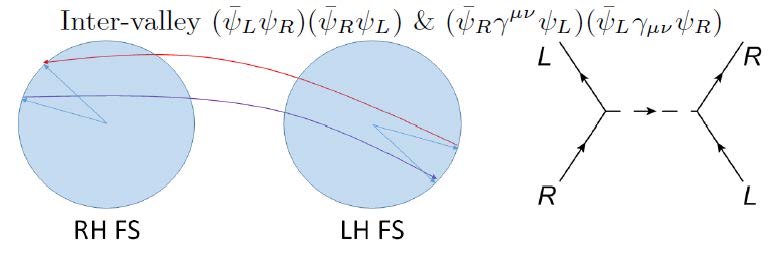}
 \caption{All possible (Lorentz-invariant) four-fermion inter-valley scattering events on the pair of chiral Fermi surfaces from (\ref{Interaction_decomposition_Inter_Valley}). } \label{inter}
\end{figure}

\begin{figure}
\includegraphics[width=8.5cm]{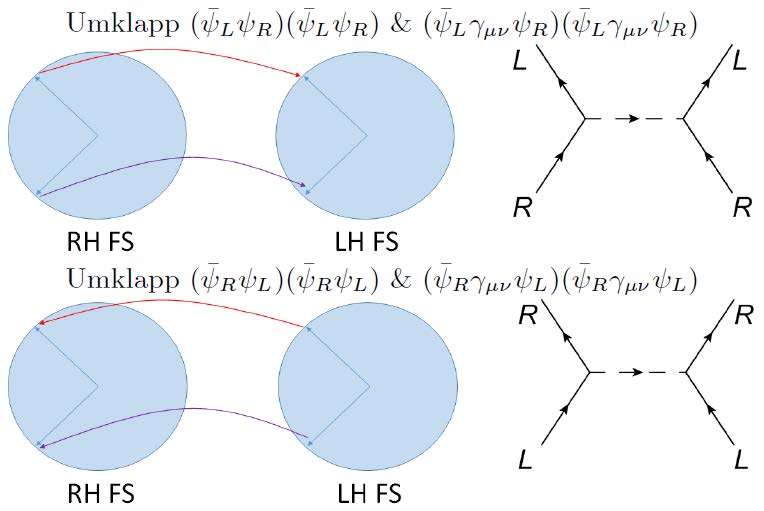}
\caption{All possible (Lorentz-invariant) four-fermion inter-valley Umklapp scattering channels on the pair of chiral Fermi surfaces from (\ref{Interaction_decomposition_Umklapp}). } \label{Umklapp}
\end{figure}

It is straightforward to rewrite these interactions in terms of chiral fermions, decomposed as follows
\bqa
\mathcal{L}_{int} &=& \mathcal{L}_{intra-valley} + \mathcal{L}_{inter-valley} + \mathcal{L}_{Umklapp} , \nn
\eqa
where intra-valley scattering events are described by
\bqa
\mathcal{L}_{intra-valley} &=& (\lambda_v + \lambda_{pv}) (\bar{\psi}_R \gamma^\mu \psi_R)(\bar{\psi}_R \gamma_\mu \psi_R) \nn
&& + (\lambda_v + \lambda_{pv}) (\bar{\psi}_L \gamma^\mu \psi_L)(\bar{\psi}_L \gamma_\mu \psi_L) \nn
&& + (2\lambda_v - 2\lambda_{pv}) (\bar{\psi}_R \gamma^\mu \psi_R)(\bar{\psi}_L \gamma_\mu \psi_L) , \nn \label{Interaction_decomposition_Intra_Valley}
\eqa
inter-valley scattering events are given by
\bqa
\mathcal{L}_{inter-valley} &=& (2 \lambda_s - 2\lambda_{ps}) (\bar{\psi}_L \psi_R)(\bar{\psi}_R \psi_L) \nn
&& + (- 2 \lambda_{as})  (\bar{\psi}_R \gamma^{\mu \nu} \psi_L)(\bar{\psi}_L \gamma_{\mu \nu} \psi_R) , \nn \label{Interaction_decomposition_Inter_Valley}
\eqa
and umklapp scattering events are expressed by
\bqa
\mathcal{L}_{Umklapp} &=&  (\lambda_s + \lambda_{ps}) (\bar{\psi}_R \psi_L)(\bar{\psi}_R \psi_L) \nn
&& + (\lambda_s + \lambda_{ps}) (\bar{\psi}_L \psi_R)(\bar{\psi}_L \psi_R) \nn
&& + \lambda_{as} (\bar{\psi}_R \gamma^{\mu \nu} \psi_L)(\bar{\psi}_R \gamma_{\mu \nu} \psi_L) \nn
&& + \lambda_{as} (\bar{\psi}_L \gamma^{\mu \nu} \psi_R)(\bar{\psi}_L \gamma_{\mu \nu} \psi_R) . \label{Interaction_decomposition_Umklapp}
\eqa
Momentum configurations in scattering events are shown in Figs. \ref{intra}, \ref{inter}, and \ref{Umklapp}, respectively. Here, we focus on only the intra- and inter-valley scattering channels, where the Umklapp channel is highly oscillating with a prefactor $ e^{\pm 4 i \bm{c} \cdot \bm{x} } $, expected to be irrelevant for generic filling.

In order to project out these four-fermion interactions into the pair of chiral Fermi surfaces, we perform a continuous angular decomposition, more convenient than using rectangular patches, given by
\bqa
\psi_\chi (x) &=& \int_{\bm{v}_\chi} \frac{d \Omega_\chi}{(2\pi)^2} \  e^{i \mu \bm{v}_\chi \cdot \bm{x} } \ e^{-i \chi \bm{c} \cdot \bm{x} } \ [q_{\chi +} ( \bm{v}_\chi, \ x ) \nn &+& q_{\chi -} ( \bm{v}_\chi, \ x )] . \label{angular_decomposition}
\eqa
Here, $\int_{\bm{v}_\chi} \frac{d \Omega_\chi}{(2\pi)^2}$ represents an angular integral for the continuous patch of $\bm{v}_\chi$. Their Fourier components are
\bqa
q_{\chi \pm} ( \bm{v}_\chi, \ x ) & = & \int_{- \infty}^{+ \infty} \frac{d \omega_{\chi}}{(2 \pi)} \int_{ \left |  \bm{l}_{\chi} \right | < \Lambda } \frac{d^3 \bm{l}_{\chi} }{(2 \pi)^3 } \nn && e^{- i \bm{l}_{\chi} \cdot \bm{x} - i \omega_{\chi} t } \ q_{\chi \pm} ( \bm{v}_\chi, \ \bm{l}_\chi, \ \omega_{\chi} ) .
\eqa
Introducing these mode decompositions into Eqs. (\ref{Interaction_decomposition_Intra_Valley}) and (\ref{Interaction_decomposition_Inter_Valley}), we find effective four-fermion interactions at low energies. Following the previous procedure, we obtain these effective interactions within the $1/\mu$ expansion, which arise from integrations of $q_{\chi -}$ fields. Considering their momentum, frequency, and field dependencies, their roles are expected to be sub-leading, compared with the $\mathcal{O}(\mu^0)$ order. In this respect we focus on effective interactions in the $\mathcal{O}(\mu^0)$ order.

Resorting to Eq. (\ref{angular_decomposition}), we represent four-fermion interactions up to the $\mathcal{O}(\mu^0)$ order as follows: The intra-valley scattering term is given by
\begin{widetext}
\bqa
\mathcal{L}_{intra-valley} = \int_{\bm{n}, l, \omega} u_{v \chi \chi'} (\bm{1},\bm{2},\bm{3},\bm{4}) [ \bar{q}_{\chi +}(\bm{3}) \gamma^\mu q_{\chi+}(\bm{1}) ] [ \bar{q}_{\chi' +}(\bm{4}) \gamma_\mu q_{\chi' +}(\bm{2}) ]  \label{Effective_Intra_Valley_Interaction}
\eqa
and the inter-valley scattering term is described by
\bqa
\mathcal{L}_{inter-valley} &=& \int_{\bm{n}, l, \omega} u_{s} (\bm{1},\bm{2},\bm{3},\bm{4}) [ \bar{q}_{\chi +}(\bm{3}) q_{\chi' +}(\bm{1}) ] [ \bar{q}_{\chi' +}(\bm{4}) q_{\chi +}(\bm{2}) ] \nn &&+ \int_{\bm{n}, l, \omega} u_{as} (\bm{1},\bm{2},\bm{3},\bm{4})  [ \bar{q}_{{\chi}+}(\bm{3}) \gamma^{\mu \nu} q_{{\chi'}+}(\bm{1}) ] [ \bar{q}_{{\chi'}+}(\bm{4}) \gamma_{\mu \nu} q_{{\chi}+}(\bm{2}) ] .  \label{Effective_Inter_Valley_Interaction}
\eqa
\end{widetext}
Here, we introduce an abbreviated notation of the chiral fermion field
\bqa && q_{\chi+} (\bm{j}) = q_{\chi+} ( \bm{n}_j, l_j, \omega_j) . \eqa
In addition, we consider the integral expression
\bqa
\int_{\bm{n}, l, \omega} & \equiv & \prod_{j = 1}^{3} \left [ \int_{\bm{n}_j} \frac{d \Omega_j}{(2\pi)^2} \int_{-\infty}^{\infty} \frac{d \omega_j}{2\pi} \int_{- \Lambda}^{\Lambda} \frac{d l_j}{2 \pi} \right ] \theta ( \Lambda - | l_4 |) , \nn
\eqa
following the Landau's Fermi-liquid theory.

It is important to notice that we have an additional kinematical dependence in the tree-level scaling, given by $\theta(\Lambda - |l_4|)$. As a result, only forward and BCS scattering channels turn out to be marginal while other interactions are all highly irrelevant, essentially the same case as the Landau's Fermi-liquid theory. Such non-vanishing four-fermion couplings are
\bqa
&& u( \bm{n}_1, \bm{n}_2, \bm{n}_3, \bm{n}_4) |_{\bm{n}_1 \cdot \bm{n}_2 = \bm{n}_3 \cdot \bm{n}_4} \nn && = F (\bm{n}_1 \cdot \bm{n}_2 \ ; \ \phi_{12;34}) = F(z, \phi)
\eqa
for forward scattering and
\bqa
u( \bm{n}_1 , - \bm{n}_1 , \bm{n}_3 , -\bm{n}_3 ) = V( \bm{n}_1 \cdot \bm{n}_3 ) = V(z_{13})
\eqa
for BCS pairing, where we follow the notation of Ref. \cite{Shankar_Review}. As a result, we rewrite Eqs. (\ref{Effective_Intra_Valley_Interaction}) and (\ref{Effective_Inter_Valley_Interaction}) in terms of forward and BCS scattering channels
\begin{widetext}
\bea
\mathcal{S}_{Forward}
&=& \int_{n, l, \omega} \Big[ \ g_{RR}^F (z, \phi) \ [ U_{\bm{3}}^{R\dagger} U^R_{\bm{1}} ]_{11} \ [ U_{\bm{4}}^{R\dagger} U^R_{\bm{2}} ]_{11} \ f_R^\dagger (\bm{3})  f_R (\bm{1}) f_R^\dagger (\bm{4}) f_R (\bm{2}) \nn
&&\hspace{30pt} + g_{LL}^F (z, \phi) \ [ U_{\bm{3}}^{L\dagger} U^L_{\bm{1}} ]_{44} \ [ U_{\bm{4}}^{L\dagger} U^L_{\bm{2}} ]_{44} \ f_L^\dagger (\bm{3}) f_L (\bm{1}) f_L^\dagger (\bm{4}) f_L (\bm{2}) \nn
&&\hspace{30pt} + g_{RL}^F (z, \phi) \ [ U_{\bm{3}}^{L\dagger} U^R_{\bm{1}} ]_{41} \ [ U_{\bm{4}}^{R\dagger} U^L_{\bm{2}} ]_{14} \ f_L^\dagger (\bm{3}) f_R (\bm{1}) f_R^\dagger (\bm{4}) f_L (\bm{2}) \Big],
\label{4fermi_Forward}
\eea
and
\bea
\mathcal{S}_{BCS}
&=& \int_{n_1 , n_3} \int_{l, \omega} \Big[ \ g_{RR}^V (z_{13}) \ [ U_{\bm{3}}^{R\dagger} U^R_{\bm{1}} ]_{11} \ [ U_{-\bm{3}}^{R\dagger} U^R_{-\bm{1}} ]_{11} \ f_R^\dagger (\bm{3}) f_R (\bm{1}) f_R^\dagger (-\bm{3}) f_R (-\bm{1}) \nn
&&\hspace{50pt} + g_{LL}^V (z_{13}) \ [ U_{\bm{3}}^{L\dagger} U^L_{\bm{1}} ]_{44} \ [ U_{-\bm{3}}^{L\dagger} U^L_{-\bm{1}} ]_{44} \ f_L^\dagger (\bm{3}) f_L (\bm{1}) f_L^\dagger (-\bm{3}) f_L (-\bm{1}) \nn
&&\hspace{50pt} + g_{RL}^V (z_{13}) \ [ U_{\bm{3}}^{L\dagger} U^R_{\bm{1}} ]_{41} \ [ U_{-\bm{3}}^{R\dagger} U^L_{-\bm{1}} ]_{14} \ f_L^\dagger (\bm{3}) f_R (\bm{1}) f_R^\dagger (-\bm{3}) f_L (-\bm{1}) \Big] ,
\label{4fermi_BCS}
\eea
\end{widetext}
respectively, after the spin enslavement [Eqs. (\ref{Spin_Enslavement_Positive_Chirality}) and (\ref{Spin_Enslavement_Negative_Chirality})]. Here, a pair of $U^\chi_{\bm p}$ matrices plays the role of an anisotropic Berry phase factor, which can be interpreted as a wave-function overlap factor. We have
\bea
g_{RR}^F ( z , \phi) & = & 4 F_{vRR} ( z_{12}, \phi_{12; 34} ), \nn
g_{LL}^F ( z, \phi) & = & 4 F_{vLL} ( z_{12}, \phi_{12; 34} ), \nn
g_{RL}^F ( z, \phi) & = & F_s ( z_{12}, \phi_{12; 34} ) - 2 F_{vRL} (z_{12}, \phi_{12;34}) \nn
\eea
for the forward scattering channel and
\bea
g_{RR}^V ( z_{13} ) & = & 4 V_{vRR} ( z_{13}), \nn
g_{LL}^V ( z_{13} ) & = & 4 V_{vLL} ( z_{13}), \nn
g_{RL}^V ( z_{13} ) & = & V_s ( z_{13} ) + 2 V_{vRL} ( z_{13} )
\eea
for the BCS channel.

\begin{figure}
\centering
 {\includegraphics[width=8.5cm]{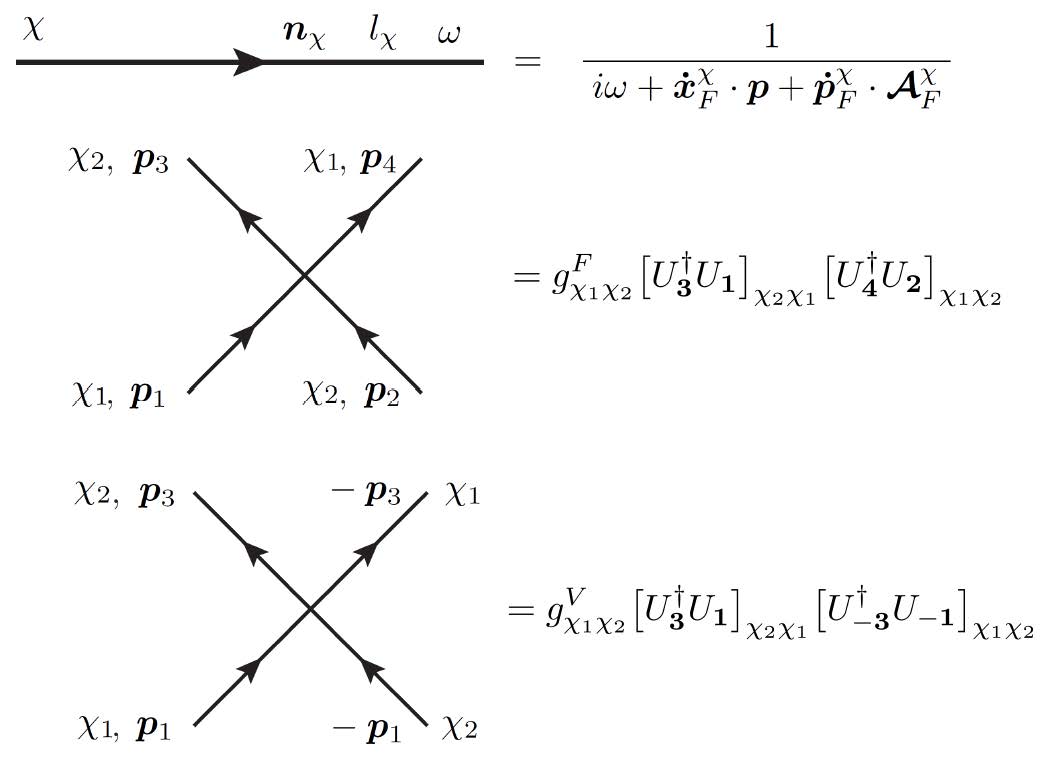}}
\caption{Feynman rules for the low energy effective action [Eq. (\ref{eq:TFGT}) in equilibrium (${\bf E}=0$)] with effective four-fermion interactions [Eqs. (\ref{4fermi_Forward}) and (\ref{4fermi_BCS})]} \label{Feynrules1}
\end{figure}

We perform the renormalization group analysis, based on the effective action for the topological Fermi-gas state [Eq. (\ref{eq:TFGT})] with effective four-fermion interactions on the pair of chiral Fermi surfaces [Eqs. (\ref{4fermi_Forward}) and (\ref{4fermi_BCS})]. See Fig. \ref{Feynrules1}, expressing Feynman rules for the perturbative renormalization group analysis.

\begin{figure}
 \centering
 {\includegraphics[width=8.5cm]{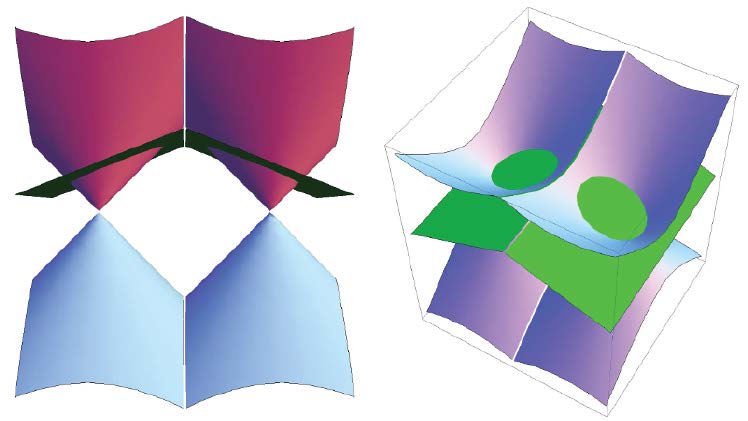}}
 \caption{A pair of chiral Fermi surfaces with the Zeeman-type shift.} \label{ellipsoid}
\end{figure}

We emphasize that dynamics of spinless fermions on the pair of chiral Fermi surfaces differs from that of conventional electrons on non-chiral Fermi surfaces, where contributions from the Zeeman-type shift and Berry-curvature effect are responsible. In particular, the Zeeman-type shift modifies the shape of a chiral Fermi surface (Fig. \ref{ellipsoid}) from a sphere to an ellipsoid. In addition, the Fermi velocity is renormalized by the Berry-curvature effect.

\begin{figure}
\includegraphics[width=9cm]{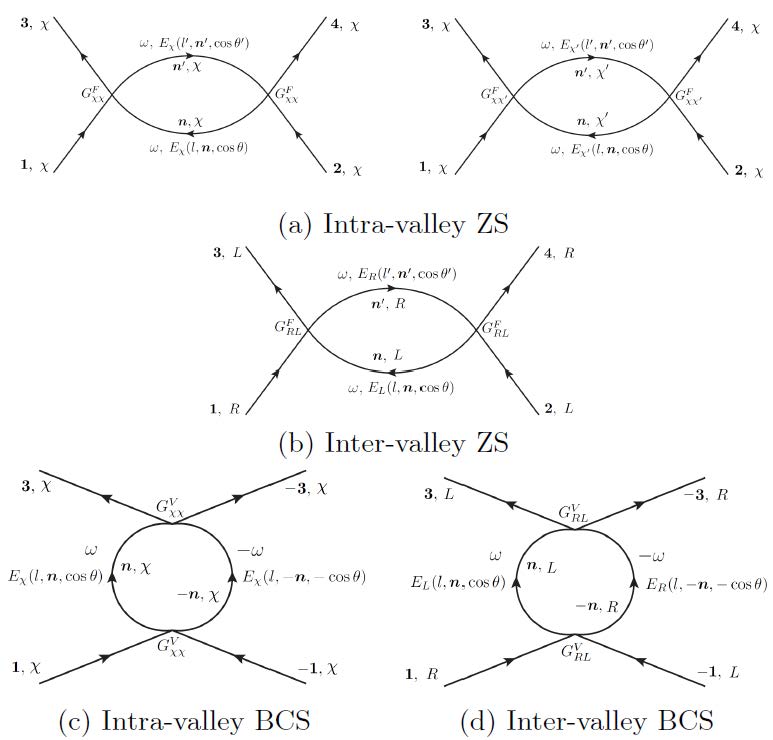}
\caption{Contributions from one-loop diagrams in the renormalization group analysis. $E_\chi (l , \ \bm{n} , \ \cos \theta)$ is the dispersion of a deformed chiral Fermi surface with chirality $\chi$.} \label{one_loop_diag}
\end{figure}

First, we focus on the forward intra-valley scattering channel, given by $g_{RR}^F$ and $g_{LL}^F$. It turns out that their renormalizations are essentially the same as those of the Landau's Fermi-liquid state except for the ellipsoid Fermi surface, where the diagrams are shown in Figs. \ref{one_loop_diag} (a) and (b), which may contribute to the renormalization group flows for $g_{RR}^F$ and $g_{LL}^F$ in the one-loop level. We note that both internal fermion lines always lie in the same chiral Fermi surface for intra-valley forward interactions.

The ZS diagrams [Fig. \ref{one_loop_diag} (a)] are given by
\bea
&& d G_{\chi \chi}^F (\bm{1},\bm{2},\bm{3},\bm{4}) \nn
&& = \int_{-\infty}^\infty \frac{d \omega}{2 \pi} \int_{d \Lambda} \frac{d l}{2 \pi} \int_{\bm{n}} \frac{d \Omega}{(2 \pi)^2} \nn
&& \frac{G_{\chi \chi}^F (\bm{n}', \bm{2}, \bm{n}, \bm{4}) \ G_{\chi \chi}^F (\bm{1}, \bm{n}, \bm{3}, \bm{n}') }{[i \omega - E_{\chi} (l, \bm{n}, \cos \theta)][i \omega - E_{\chi} (l', \bm{n}', \cos \theta ') ]} \nn
&& + \int_{-\infty}^\infty \frac{d \omega}{2 \pi} \int_{d \Lambda} \frac{d l}{2 \pi} \int_{\bm{n}} \frac{d \Omega}{(2 \pi)^2} \nn
&& \frac{G_{\chi \chi'}^F (\bm{2}, \bm{n}', \bm{4}, \bm{n}) \ G_{\chi \chi'}^F (\bm{1}, \bm{n}, \bm{n}', \bm{3}) }{[i \omega - E_{\chi'} (l, \bm{n}, \cos \theta)][i \omega - E_{\chi'} (l', \bm{n}', \cos \theta ') ]} \ , \label{ZS_ZS2_Forward_intra_RG} \nn
\eea
where $ G_{\chi \chi'}^F (\bm{1},\bm{2},\bm{3},\bm{4}) \equiv g_{\chi \chi'}^F [U^{\chi'\dagger}_{\bm{3}} U^\chi_{\bm{1}}] [U^{\chi\dagger}_{\bm{4}} U^{\chi'}_{\bm{2}}]$ is a vertex factor with Berry phase. We note $ \chi \neq \chi' $ in the second line. $ E_{\chi} (l, \bm{n}, \cos \theta) $ is a modified dispersion on an elliptic chiral Fermi surface with chirality $\chi$. It is straightforward to see that the frequency integral vanishes identically since both poles in the pair of fermion Green's functions reside in the same side of the frequency space.

\begin{figure}
\centering \includegraphics[width=7cm]{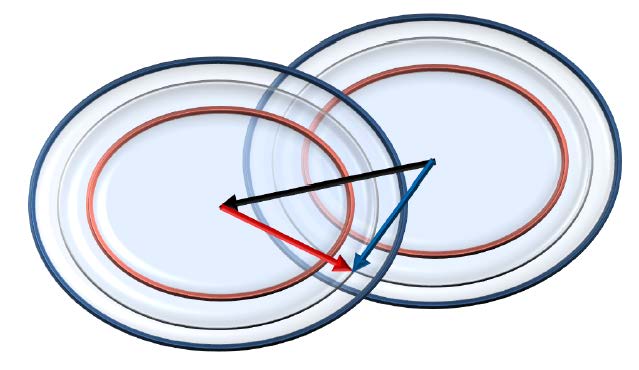}
\caption{Available phase space for the forward scattering channel. Red and blue arrows represent internal momenta in one-loop diagrams of the renormalization group analysis for forward interactions. Phase space overlap occurs only in the intersection between red and blue lines with $d\Lambda$ thickness.} \label{RGforwardChannelPhaseSpaceRestriction}
\end{figure}

There are also ZS' and BCS diagram contributions. The ZS' diagram can be constructed in the same way as the above, given by $\bm{3} \leftrightarrow \bm{4}$ in Eq. (\ref{ZS_ZS2_Forward_intra_RG}). The BCS diagram is also given similarly. It turns out that they give rise to the order of $\mathcal{O}(d \Lambda^2 )$, proportional to the phase space volume of intersections between red and blue thin bars in Fig. \ref{RGforwardChannelPhaseSpaceRestriction}.

\begin{figure}
 \centering \includegraphics[width=7cm]{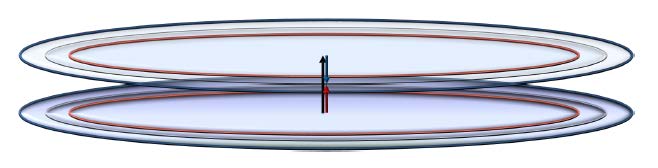}
\caption{Available phase space for the forward scattering channel. Red and blue arrows represent internal momenta in one-loop diagrams of the renormalization group analysis for forward interactions. Phase space overlap occurs only in the intersection between red and blue lines with $d\Lambda$ thickness. When the dispersion relation is distorted by strong magnetic fields seriously, the intersection for the exchange scattering channel seems to have the volume of $\mathcal{O}(d\Lambda)$. However, considering the azimuthal angle, the overall overlap volume turns out to be of order $\mathcal{O}(d\Lambda^2 )$.
} \label{Phase_Space_Diagram2}
\end{figure}

Although there exist some nested line-pair configurations in the intra-valley channel (Fig. \ref{Phase_Space_Diagram2}), their bare couplings are highly suppressed or exactly vanish due to the Berry phase factor. As a result, forward intra-valley interactions given by $g_{RR}^F$ and $g_{LL}^F$ are marginal up to the one-loop level.

\begin{figure*}
\centering
{\includegraphics[width=14cm]{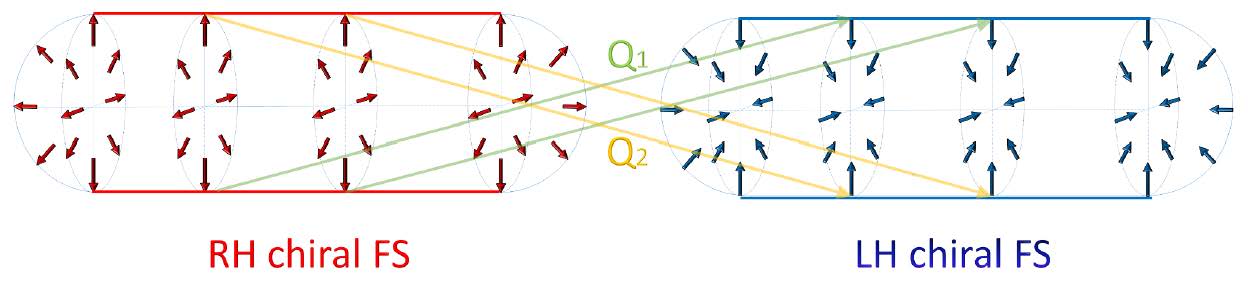}}
\caption{Nested line pairs in the inter-valley channel. This situation is realized under strong magnetic fields, which distort the Fermi surface along the direction of the applied magnetic field. Red and blue arrows denote the $CP^1$ Berry phase factor given by $U_{\bm{p}}$ (spin orientations). A pair of red and blue lines form a nesting pair connected by the nesting vector ${\bf Q}_1$ or ${\bf Q}_2$. There are infinitely many nesting pairs due to the azimuthal symmetry in these Fermi surfaces. Notice that two red lines or two blue lines are also nested to each other. However, the Berry-phase factor kills their bare couplings completely due to the wave-function orthogonality.} \label{nestinglinepair}
\end{figure*}

Second, we consider the inter-valley scattering channel, given by $g_{RL}^F$, where two internal momenta in diagrams lie on the opposite chiral Fermi surfaces. The ZS diagram [Fig. \ref{one_loop_diag} (b)] is given by
\bea
&&d G_{RL}^F (\bm{1},\bm{2},\bm{3},\bm{4}) = \int_{-\infty}^\infty \frac{d \omega}{2 \pi} \int_{d \Lambda} \frac{d l}{2 \pi} \int_{\bm{n}} \frac{d \Omega}{(2 \pi)^2} \nn
&& \frac{G_{RL}^F (\bm{n}', \bm{2}, \bm{n}, \bm{4}) \ G_{RL}^F (\bm{1}, \bm{n}, \bm{3}, \bm{n}') }{[i \omega - E_{L} (l, \bm{n}, \cos \theta)][i \omega - E_{R} (l', \bm{n}', \cos \theta ') ]} \nn
\eea
The ZS' diagram can be found from $\bm{3} \leftrightarrow \bm{4}$ in the above equation, and the BCS is constructed similarly. It turns out that the overlap region is the same as that of the intra-valley case. In most cases, the momentum integral is restricted within a small intersection volume of order $\mathcal{O}(d \Lambda^2 )$ as before. But, there are some interesting channels between two nested lines, expected to show enhancement of an overlap volume in the phase space along the elongated axis of the ellipsoid (Fig. \ref{Phase_Space_Diagram2}). These nested line pairs are indicated in Fig. \ref{nestinglinepair}.

As shown in Fig. \ref{Phase_Space_Diagram2}, the intersection seems to be no longer a small rectangular region of order $\mathcal{O}(d \Lambda^2 )$. Instead, it looks to be of order $\mathcal{O}(d \Lambda )$. However, a careful analysis confirms that the intersection region is a thin-line-type, not a thin-plane-type, where the phase space volume is $\sim k_F (d \Lambda)^2$ instead of $\sim k_F^2 d \Lambda$. As a result, the forward inter-valley channel $g_{RL}^F$ is also marginal up to the one-loop level.

We find one-loop renormalizations for pairing interactions of $g_{RR}^V$, $g_{LL}^V$, and $g_{RL}^V$ [Figs. \ref{one_loop_diag} (c) and (d)], given by
\bqa
&& d G_{\chi \chi}^V (\bm{1},-\bm{1},\bm{3},-\bm{3}) \nn
&& = \int_{-\infty}^\infty \frac{d \omega}{2 \pi} \int_{d \Lambda} \frac{d l}{2 \pi} \int_{\bm{n}} \frac{d \Omega}{(2 \pi)^2} \nn
&& \frac{G_{\chi \chi}^V (\bm{n}, -\bm{n}, \bm{3}, -\bm{3}) \ G_{\chi \chi}^V (\bm{1}, -\bm{1}, \bm{n}, -\bm{n}) }{[i \omega - E_{\chi} (l, \bm{n}, \cos \theta)][- i \omega - E_{\chi} (l, -\bm{n}, -\cos \theta) ]} \nn
\label{BCS_intra_RG}
\eqa
and
\bqa
&& d G_{R L}^V (\bm{1},-\bm{1},\bm{3},-\bm{3}) \nn
&& = \int_{-\infty}^\infty \frac{d \omega}{2 \pi} \int_{d \Lambda} \frac{d l}{2 \pi} \int_{\bm{n}} \frac{d \Omega}{(2 \pi)^2} \nn
&& \frac{G_{RL}^V (-\bm{n}, \bm{n}, \bm{3}, -\bm{3}) \ G_{RL}^V (\bm{1}, -\bm{1}, \bm{n}, -\bm{n}) }{[i \omega - E_{L} (l, \bm{n}, \cos \theta)][- i \omega - E_{R} (l, -\bm{n}, -\cos \theta) ]} , \nn
\label{BCS_inter_RG}
\eqa
respectively. In Eq. (\ref{BCS_intra_RG}), Cooper pairs occur from the same chiral Fermi surface. On the other hand, Cooper pairs result from both chiral Fermi surfaces in Eq. (\ref{BCS_inter_RG}). An interesting feature may come from the presence of the Berry phase. Since we focus on the normal state above a certain temperature scale involved with the instability of the particle-particle channel and below some critical interaction parameters associated with various particle-hole instabilities, we do not consider these superconducting instabilities any more in the present study.

The above discussion allows us to identify Landau's interaction parameters in the energy functional of Eq. (\ref{Energy_Functional_TFL}) as follows
\bea
F_{\chi \chi} (\bm{p}, \bm{p}') & = & g_{\chi \chi}^F (z, 0), \nn
F_{RL} (\bm{p}, \bm{p}') & = & \big| [U^{R\dagger}_{\bm{p}} U^L_{\bm{p'}}]_{14} \big|^2 g_{RL}^F (z, 0) .
\eea
This completes the derivation of the topological Fermi-liquid theory for interacting Weyl metals from an effective microscopic model Eq. (\ref{TFL_Dirac_Theory}) [or Eq. (\ref{TFL_Dirac_Theory_Interactions})].

\section{Discussion}

Topological Fermi-gas theory should be distinguished from Landau's Fermi-liquid theory with Berry curvature, where electromagnetic properties of the former state are described by axion electrodynamics while those of the latter are governed by Maxwell electrodynamics. In order to understand the Berry curvature and the chiral anomaly more deeply, we repeat exactly the same analysis as that of the previous section but for the theory of three dimensional (one time and two space dimensions) quantum electrodynamics (QED$_{3}$), which may be regarded to be realized in graphene.

\begin{figure}
\includegraphics[width=8.5cm]{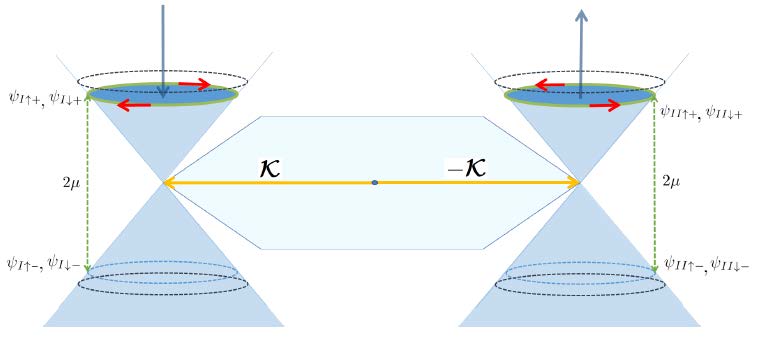}
\caption{Spectrum of graphene. $I = \bm{\mathcal{K}}$ and $II = -\bm{\mathcal{K}}$ are degeneracy points (valleys) with opposite chiralities in the first Brillouin zone, where the Berry connection $\bm{\mathcal{A}}_a$ and and associated Berry curvature $\mathcal{B}_a$ are expressed by red and blue arrows, respectively.} \label{21masslessspectrumberrycurv}
\end{figure}

First, we consider the massless case, shown in Fig. \ref{21masslessspectrumberrycurv} and described by
\bqa &&
\mathcal{L} = \sum_{a = I, II} \sum_{s = \uparrow, \downarrow} \bar{\psi}_{a s} (\gamma_a^\mu i D_\mu + \mu \gamma_a^0) \psi_{a s} .
\eqa
$\psi_{a s}$ represent two-component spinors with a valley index $a = I, ~ II$ ($I = \bm{\mathcal{K}}$ and $II = - \bm{\mathcal{K}}$) and a spin index $s = \uparrow, ~ \downarrow$, where the chirality of the $\bm{\mathcal{K}}$ valley is opposite to that of $- \bm{\mathcal{K}}$ and the spin degeneracy may be regarded as just replicas. The Dirac matrix of $\gamma_a^\mu = ( \tau^z, -i\tau^x, \chi i \tau^y )$ at each valley are given by the Pauli's matrix $\tau^i$, which acts on the sublattice space. $\chi = +1 ~ (-1)$ represents the chirality for $I = \bm{\mathcal{K}} ~ (II = - \bm{\mathcal{K}})$. In this case we find the Berry gauge field and associated Berry curvature field as
\bqa &&
\bm{\mathcal{A}}_a = - \frac{\chi}{2\mu} \bm{\hat{\phi}}_{a}
\eqa
and
\bqa
\mathcal{B}_a = (\nabla_{\bm{p}_a} \times \bm{\mathcal{A}}_a)_z = - \delta (\bm{p}_a) \frac{\chi}{2\mu^2} ,
\eqa
respectively, which corresponds to a vortex ($\chi = + 1$) and anti-vortex ($\chi = - 1$) configuration in the momentum space, where $\bm{\hat{\phi}}_{a}$ is an angular unit vector at the Fermi surface of each valley. This vortex and anti-vortex configuration will not give any Berry curvature effects on the dynamics of these Dirac electrons near the pair of Fermi surfaces. Following the same procedure before, we obtain
\bea
\mathcal{L}
&=& \mathcal{L}_0 + \mathcal{L}_{-1} + \mathcal{L}_{-2} + \mathcal{O}(\mu^{-3}), \nn
\mathcal{L}_0
&=& \sum_{a=I,II} \sum_{\bm{v}_a} \ q_{a+}^\dagger ( V_a^{\mu} i D_{\mu} ) q_{a+}, \nn
\mathcal{L}_{-1}
&=& \sum_{a=I,II} \sum_{\bm{v}_a} \frac{1}{2\mu} q_{a+}^\dagger (iD_{\perp a})^2 q_{a+}, \nn
\mathcal{L}_{-2}
&=& - \sum_{a=I,II} \sum_{\bm{v}_a} \frac{1}{4\mu^2} q_{a+}^\dagger [ i (\bm{v}_a \times\bm{E}) \cdot (\bm{v}_a \times i \bm{D}) ] q_{a+} \nn
&& - \sum_{a=I,II} \sum_{\bm{v}_a} \frac{1}{4\mu^2} \chi q_{a+}^\dagger \tau^z [ (\bm{E}\times i\bm{D})_z \nn
&& - (\bm{E}\times\bm{v}_a)_z (\bm{v}_a \cdot i\bm{D}) - (\bm{v}_a \cdot \bm{E})(\bm{v}_a \times i\bm{D})_z  ] q_{a+} \ , \label{massless21inthighE} \nn
\eea
where the $q_{a+}$ field describes the dynamics of Dirac electrons on the pair of Fermi surfaces with the vortex and anti-vortex pair configuration. All symbols are defined similarly with those of Eqs. (\ref{TFL_Zeroth_Order}), (\ref{TFL_First_Order}), and (\ref{TFL_Second_Order}). Although Berry curvature involved (chirality dependent) contributions do appear formally, the presence of $\tau^{z}$ in the last line of $\mathcal{L}_{-2}$ does not allow their net effects on transport phenomena, which may be identified with Landau's Fermi-liquid state.
%
%
%
%

%
%
\begin{figure}
\includegraphics[width=8.5cm]{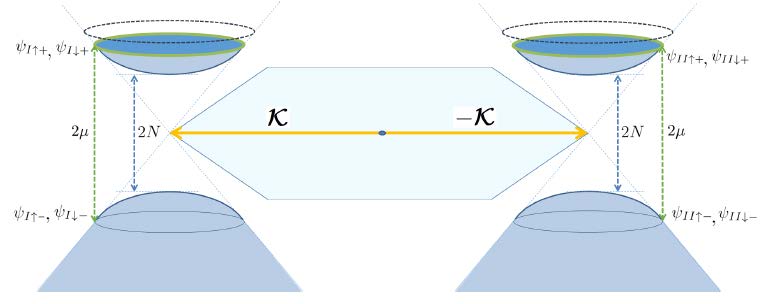}
\caption{Spectrum of graphene but with ``chiral" symmetry breaking.} \label{21massivespectrum}
\end{figure}
%
%

Second, we consider the massive case, shown in Fig. \ref{21massivespectrum} and described by
\bqa
\mathcal{L} = \sum_{a=I,II} \sum_{s=\uparrow,\downarrow} \bar{\psi}_{as} (\gamma_a^\mu i D_\mu + \mu \gamma_a^0 - N_s) \psi_{as} , \eqa
where the mass gap is assumed to result from antiferromagnetic correlations, thus $N_\uparrow = N$ and $N_\downarrow = -N$.

\begin{figure}
\centering \includegraphics[width=8.5cm]{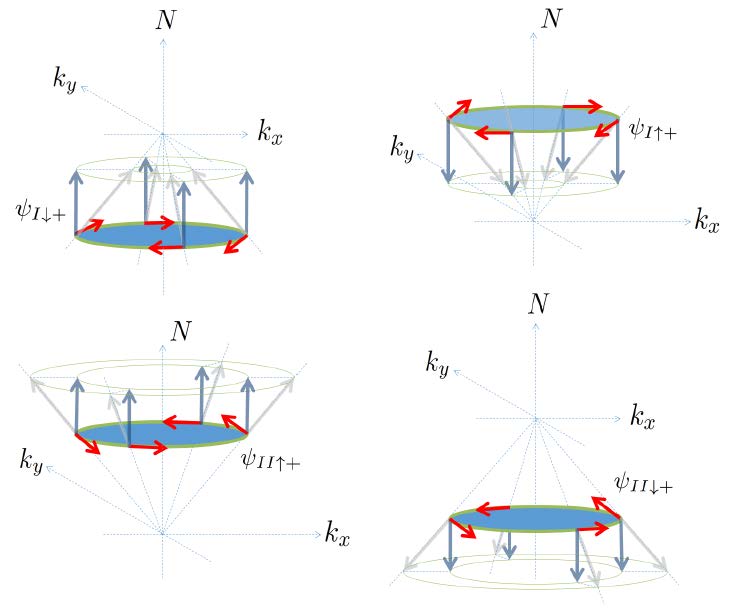}
\caption{Berry connections and curvatures in the parameter space of $(k_x, k_y, N_s)$, represented by red arrows and blue arrows, respectively.} \label{21massiveberrycurv}
\end{figure}

In this case the Berry connection and Berry curvature are given by
\bqa
\bm{\mathcal{A}}_{as}
&=& - \frac{\chi}{k_F} \sin^2 (\theta_s /2) \ \bm{\hat{\phi}}_{a}
\eqa
and
\bqa
\mathcal{B}_{as}
&=& (\nabla_{\bm{p}_a} \times \bm{\mathcal{A}}_{as})_z = - \chi \frac{N_s}{2\mu^3},
\eqa
which corresponds to a magnetic monopole and anti-monopole configuration in the parameter space (or a projected monopole and anti-monopole configuration in the momentum space), where $\cos \theta_s \equiv N_s / \mu $ and $ \mu^2 = k_F^2 + N_s^2$. See Fig. \ref{21massiveberrycurv}.

%
%
%
%

Then, we find
\bea
\mathcal{L}
&=& \mathcal{L}_0 + \mathcal{L}_{-1} + \mathcal{L}_{-2} + \mathcal{O}(\mu^{-3}) \ , \nn
\mathcal{L}_0
&=& \sum_{a=I,II} \sum_{s=\uparrow, \downarrow} \sum_{\bm{v}_a} \ q_{as+}^\dagger ( V_a^{\mu} i D_{\mu} ) q_{as+} \ , \nn
\mathcal{L}_{-1}
&=& \sum_{a=I,II} \sum_{s=\uparrow, \downarrow} \sum_{\bm{v}_a} \frac{1}{2\mu} q_{as+}^\dagger [(iD_{\perp a})^2 - \chi (\hat{\bm{K}}_s)_z B ] q_{as+} , \nn
\mathcal{L}_{-2}
&=& - \sum_{a=I,II} \sum_{s=\uparrow, \downarrow} \sum_{\bm{v}_a} \frac{1}{4\mu^2} q_{as+}^\dagger [ i (\bm{v}_a \times\bm{E}) \cdot (\bm{v}_a \times i \bm{D}) \nn
&& + \chi (\hat{\bm{K}}_s)_z (\bm{E} \times i\bm{D})_z ] q_{as+} ,
\eea
where $\bm{K}_s = ( k_F \cos \phi , \ k_F \sin \phi , \ N_s ) $, $\bm{\alpha}_a = (\gamma_a^0 \gamma_a^1 , \ \gamma_a^0 \gamma_a^2 , \ \gamma_a^0) $, and $\bm{D} = (D_x, \ D_y, \ 0) $. If one may point out that essentially the same low-energy effective action has been found except for the dimensionality, the observation is completely correct. However, this explicit demonstration proves that only the Berry-curvature effect such as the anomalous Hall effect is allowed in two dimensions while the axion electrodynamics involved with the chiral anomaly cannot occur, which originates from the constraint involved with the dimensionality. Here, even the anomalous Hall effect is not observed since the anomalous Hall effect from each valley turns out to be canceled.

%
%
\begin{table*}
\caption{Classification of low-energy effective field theories in the presence of Berry curvature in $d=2$ and $d=3$. UV and IR represent ultraviolet and infrared, respectively. LFL and TFL express Landau's Fermi-liquid and topological Fermi-liquid, respectively. AHE is the abbreviation of anomalous Hall effect.}
\centering
\begin{tabular}{ c | c | c | c}
\hline\hline
\multirow{2}*[-1.5ex]{\ \ \ UV Field Theory \ \ \ } & \multirow{2}*[-0.3ex]{QED$_{3}$ with} & \multirow{2}*[-0.3ex]{QED$_{3}$ with} & \multirow{2}*[-0.3ex]{QED$_{4}$ with} \\
 & \multirow{2}*[-0.3ex]{Massless Spectrum} & \multirow{2}*[-0.3ex]{Massive Spectrum} & \multirow{2}*[-0.3ex]{Inhomogeneous Axion Term} \\ 
 & & & \\
\hline
 & \multicolumn{3}{c}{} \\
IR Field Theory & \multicolumn{3}{c}{$\ \mathcal{S} = \int_{0}^{\beta}\! d \tau \int\! d^{d} \bm{x} \ f_{\chi}^{\dagger}(\bm{x},\tau) \Big( \partial_{\tau} - \varphi + i  G_d^{\chi -1} \bm{\dot{x}}_{F}^{\chi} \cdot [\bm{\nabla}_{\bm{x}} + i \bm{A} ] - \boldsymbol{\dot{p}}_{F}^{\chi} \cdot \bm{\mathcal{A}}_{F}^{\chi} \Big) f_{\chi}(\bm{x},\tau)$} \\ 
 & \multicolumn{3}{c}{} \\ \cline{2-4}
\hline
 & & & \\
 & $\dot{x}_i = v_i $ \hspace{40pt} & \ \ \ $\dot{x}_i^\chi = G_2^\chi [v_i + \epsilon^{ij} E_j \mathcal{B}^\chi]$ \ \ \ & \ \ \ $\dot{\bm{x}}^\chi = G_3^\chi [\bm{v}_{\bm{p}} + \bm{E} \times \bm{\mathcal{B}}_{\bm{p}}^\chi + \bm{B} (\bm{v}_{\bm{p}} \cdot \bm{\mathcal{B}}_{\bm{p}}^\chi)]$ \ \ \ \\
 & & & \\
 & \ \ \ $\dot{p}_i = E_i + \epsilon^{ij} v_j B$ \ \ \ & $\dot{p}_i^\chi = G_2^\chi [E_i + \epsilon^{ij} v_j B]$ \ &  $\dot{\bm{p}}^\chi = G_3^\chi [\bm{E} + \bm{v}_{\bm{p}} \times \bm{B} +\bm{\mathcal{B}}_{\bm{p}}^\chi (\bm{E} \cdot \bm{B})]$ \ \\ 
 & & & \\
\cline{2-4}
 & \multirow{2}*[-.3ex]{Drude model} & \multirow{2}*[-.3ex]{Drude model with} & \multirow{2}*[-1.5ex]{``Topological'' Drude model} \\
 & \multirow{2}*[-.3ex]{(Plain vanilla)} & \multirow{2}*[-.3ex]{Berry curvature} & \multirow{2}*[-.3ex]{} \\
 & & & \\
\hline
\multirow{2}*[-1.5ex]{} & \multirow{2}*[-.3ex]{$\mathcal{B}_a = - \delta (\bm{p}_a) \frac{\chi}{2\mu^2}$} & \multirow{2}*[-.3ex]{$\mathcal{B}_{as} = - \chi \frac{N_s}{2\mu^3}$} & \multirow{2}*[-.3ex]{$\bm{\mathcal{B}}^\chi = \chi \frac{1}{2\mu^2} \hat{\bm{p}} $} \\
\multirow{2}*[0ex]{Berry Curvature} &  &  & \\ \cline{2-4}
 & \multirow{2}*[-.3ex]{Vortex} & \multirow{2}*[-.3ex]{Projected monopole} & \multirow{2}*[-.3ex]{Monopole} \\ 
 & & &\\
\hline
 & \multicolumn{3}{c}{} \\
Transport Theory & \multicolumn{3}{c}{$\Bigl( \frac{\partial}{\partial T} + \bm{\dot{x}}_{F}^{\chi} \cdot \bm{\nabla}_{\bm{R}} + \boldsymbol{\dot{p}}_{F}^{\chi} \cdot \bm{\nabla}_{\bm{p}} \Bigr) G_{\chi}^{<}(\bm{p};\bm{R},T) = 0$} \\
 & \multicolumn{3}{c}{} \\ \cline{2-4}
\hline
\multirow{2}{*}{Response} & \multirow{2}{*}{-} & \multirow{2}{*}{AHE} & \multirow{2}{*}{AHE \& Axion electrodynamics} \\
 & & & \\
\hline
\multirow{2}*[-1.5ex]{\ \ Four-Fermion Interactions \ \ } & \multirow{2}*[-1.5ex]{LFL} & \multirow{2}*[-.3ex]{LFL with} & \multirow{2}*[-1.5ex]{TFL} \\
\multirow{2}*[-.3ex]{} & & \multirow{2}*[-.3ex]{Berry curvature} & \\ 
 & & & \\
\hline\hline
\end{tabular}
\label{classification of effective actions}
\end{table*}
%
%

The above discussion leads us to construct table \ref{classification of effective actions} as our conclusion. First of all, the low-energy effective field theory turns out to be the same as each other completely. On the other hand, governing equations of motion are shown to differ, identified with Drude model, Drude model with Berry curvature, and topological Drude model and responsible for classifying liquids into Landau's Fermi liquid, Landau's Fermi liquid with Berry curvature, and topological Fermi liquid, respectively. Although both the Landau's Fermi liquid and Landau's Fermi liquid with Berry curvature may be classified into the same Landau's Fermi-liquid state described by Landau's Fermi-liquid theory, where their electromagnetic properties are described by the Maxwell electrodynamics, the topological Fermi-liquid state should be regarded to differ in the respect that there exists a topologically protected surface Fermi-arc state.

\section{Conclusion}

Generally speaking, one may neglect high-energy electron excitations deep inside a Fermi surface if there do not exist a magnetic monopole and anti-monopole pair in momentum space. However, if the Fermi surface encloses a magnetic monopole, we must take into account the role of high-energy excitations of electrons in the renormalized effective field theory, where the topological information such as Berry curvature and chiral anomaly are incorporated through these high-energy excitations. Indeed, we could observe the evidence in the first quantization, where the utilization of the diagonal basis takes the topological structure. We did essentially the same work in the second quantization, where high-energy electronic fluctuations turn out to be responsible for the topological information in the low-energy dynamics of chiral fermions, described by the topological Fermi-liquid theory.

The nature of such spinless fermions in the topological Fermi-liquid theory differs from that of electron quasiparticles in the Landau's Fermi-liquid theory. Actually, such fermions carry not only the electric quantum number described by the minimal gauge-field coupling term but also the magnetic dipole moment assigned from the Berry curvature. The current of chiral spinless fermions couples to the external magnetic field directly, given by $i \bm{\dot{x}}_{F}^{\chi} \cdot f_{\chi}^{\dagger}(\bm{r},\tau) (\bm{\nabla}_{\bm{r}} + i \bm{A}) f_{\chi}(\bm{r},\tau)$, where the renormalized velocity is given by $\bm{\dot{x}}_{F}^{\chi} = \bm{\dot{x}}_{F}^{\chi}[\bm{E},\bm{B};\bm{\mathcal{B}}_{F}]$, Eq. (\ref{Drude_Model_Group_Velocity}).

In summary, we derived a topological Fermi-liquid theory from QED$_{4}$ with a chiral gauge field, describing a topological Fermi-liquid fixed point for a pair of chiral Fermi surfaces identified with an interacting Weyl metallic state. We emphasize that this effective theory differs from the Landau's Fermi-liquid theory, describing the Landau's Fermi-liquid fixed point, although the concept of electron quasiparticles remains valid at this topological Fermi-liquid fixed point.

Recently, we found that the $B^{2}$ enhancement is not limited on the longitudinal magnetoconductivity \cite{Boltzmann_Chiral_Anomaly5}. Both the Seebeck and thermal conductivities in the longitudinal setup have been predicted to show essentially the same enhancement proportional to $B^{2}$. Most surprisingly, a topologically modified Boltzmann transport theory with both the Berry curvature and chiral anomaly has predicted that the Wiedemann-Franz law is violated only in the longitudinal setup, showing the $B^{2}$ dependence in the Lorentz number, which turns out to be purely topological, more precisely, geometrical in the origin. Since the breakdown of the Wiedemann-Franz law appears in spite of the existence of electron quasiparticles, the Weyl metallic state cannot be identified with the Landau's Fermi-liquid fixed point.

The topological Fermi-liquid theory serves a theoretical platform for us to investigate the role of Fermi-liquid interactions in anomalous transport phenomena of interacting Weyl metals such as anomalous Hall effects, chiral magnetic and vortical effects, and negative longitudinal magnetoresistivity properties. In addition, it allows us to study how thermodynamic properties such as the Wilson's ratio and spectra of collective excitations such as zero sound modes in the Landau's Fermi-liquid state are modified due to the Berry curvature and the chiral anomaly. Furthermore, symmetry breaking phase transitions from the topological Fermi-liquid state would be described by a topologically modified Landau-Ginzburg-Wilson theory. We speculate that some types of topological-in-origin terms may arise to allow nontrivial quantum numbers in topological excitations of local order parameters involved with symmetry breaking.

\section*{Acknowledgement}

This study was supported by the Ministry of Education, Science, and Technology (No. NRF-2015R1C1A1A01051629 and No. 2011-0030046) of the National Research Foundation of Korea (NRF) and by TJ Park Science Fellowship of the POSCO TJ Park Foundation. This work was also supported by the POSTECH Basic Science Research Institute Grant (2016). We would like to appreciate fruitful discussions in the APCTP Focus program ``Lecture Series on Beyond Landau Fermi Liquid and BCS Superconductivity near Quantum Criticality" (2016). KS appreciates fruitful discussions and collaborations with experimentalists of Heon-Jung Kim, Jeehoon Kim, and M. Sasaki.

\end{document}